\documentclass[10pt,twocolumn]{article}

\pdfoutput=1

\usepackage{url}
\usepackage{graphicx}
\usepackage{amsmath}
\usepackage{amsfonts}
\usepackage{amssymb}
\usepackage{color}

\begin{document}
\title{LBIBCell: A Cell-Based Simulation Environment for Morphogenetic Problems}
\author{Simon Tanaka$^{1,2\footnote{simon.tanaka@bsse.ethz.ch}}$, David Sichau$^{1}$, and Dagmar Iber$^{1,2\footnote{dagmar.iber@bsse.ethz.ch}}$}

\maketitle

\begin{abstract}

\section{Motivation:}
The simulation of morphogenetic problems requires the simultaneous and coupled simulation of signalling and tissue dynamics. A cellular resolution of the tissue domain is important to adequately describe the impact of cell-based events, such as cell division, cell-cell interactions, and spatially restricted signalling events. A tightly coupled cell-based mechano-regulatory simulation tool is therefore required.

\section{Results:}
We developed an open-source software framework for morphogenetic problems. The environment offers core functionalities for the tissue and signalling models. In addition, the software offers great flexibility to add custom extensions and biologically motivated processes. Cells are represented as highly resolved, massless elastic polygons; the viscous properties of the tissue are modelled by a Newtonian fluid. The Immersed Boundary method is used to model the interaction between the viscous and elastic properties of the cells, thus extending on the IBCell model. The fluid and signalling processes are solved using the Lattice Boltzmann method. As application examples we simulate signalling-dependent tissue dynamics. 
\section{Availability:}
The documentation and source code are available on {\url{http://tanakas.bitbucket.org/lbibcell/index.html}}

\section{Contact:} dagmar.iber@bsse.ethz.ch
\end{abstract}

\section{Introduction}

During morphogenesis, tissue grows and self-organises into complex functional units such as organs.
The process is tightly controlled, both by signalling and by mechanical interactions.
Long-range signalling interactions in the tissues can be mediated by diffusible substances, called morphogens, and by long-range cell processes (\cite{Restrepo2014}).
The dynamics of the diffusible factors can typically be well described by systems of continuous reaction-advection-diffusion partial differential equations (PDE).
The appropriate tissue representation depends on the relevant time scale. For a homogeneous isotropic embryonic tissue, experiments show that the tissue is well approximated by a viscous fluid on long time scales (equilibration after 30 minutes to several hours) and by an elastic material on short time scales (seconds to minutes) (\cite{Forgacs1998}).
However, biological control typically happens on a shorter time scale, and many cellular processes such as cell migration and adhesion, cell polarity, directed division, monolayer structures and differentiation cannot be cast into a continuous formulation in a straight-forward way.
A number of cell based simulation techniques at different scales and different level of detail have been developed to study these processes; here, we discuss main representatives for each category.

The \textit{Cellular Potts model}, introduced by \cite{Graner1992}, is solved on a lattice, with each lattice point holding a generalized spin value denoting cell identity.
Similar to the Ising model, Hamiltonian energy functions are formulated and minimized using a Metropolis algorithm.
It has been applied to a multitude of problems, and is implemented in the software \textit{CompuCell3D} (\cite{Swat2012}).
However, the correspondence between the biological problem and the Hamiltonian, the temperature and the time step is not always straightforward.

The \textit{subcellular element model} divides cells into subcellular elements, which are represented by computational particles.
The elements interact via interacting potentials which are subject to modelling.
The motion of the elements is governed by overdamped Langevin dynamics,
such that the method is mesh-free.
The framework was first introduced by \cite{Newman2005} and later applied by \cite{Sandersius2011, Sandersius2011a}.
This approach allows for detailed biophysical modelling, both in 2D and 3D.

The \textit{spheroid model} developed by \cite{Drasdo2007} assumes that cells in unstructured cell populations are similar to colloidal particles.
The cells are modelled as point particles, hosting interaction potentials.
Their motion consist of a random and a directed movement.
Neighboring cells form adhesive bonds, which are represented using models borrowed from contact mechanics, such as e.g.
the Johnson-Kendall-Roberts model (\cite{Chu2005}).
Many cellular processes such as cell shape change, division, death, lysis, cell-cell interaction and migration have
been successfully translated into the spheroid model (\cite{Drasdo2007}).
Intra- and extracellular diffusion has not yet been introduced and implemented.
The spheroid model extends efficiently to 3D, and it has been implemented in the open-source framework \textit{CellSys} (\cite{Hoehme2010a}).

The \textit{vertex model} uses polygons (or polyhedra in 3D) to represent cells in densely packed tissues, e.g. in \textit{Drosophila} wing disc epithelia (\cite{Farhadifar2007}).
For each vertex, forces are computed - either via a potential or directly.
The vertices are moved subsequently according to overdamped equations of motion, or via a Monte Carlo algorithm.
The model is implemented in the open-source software \textit{Chaste} (\cite{Pitt-Francis2009}).

The \textit{viscoelastic cell model} (also called IBCell models) presented in \cite{Rejniak2004} and \cite{Rejniak2007} uses the immersed boundary method (\cite{Peskin2003}) to represent individually deformable cells as immersed elastic bodies.
The cytoplasm and the extracellular matrix and fluid are represented by a viscous incompressible
fluid.
In this framework, a vast amount of biological processes such as cell growth, cell division, apoptosis and polarization has been realized.
The model was applied to study tumor and epithelial dynamics.
Due to the very high level of detail, the viscoelastic cell model is computationally expensive and has not yet been implemented in 3D.

The software framework \textit{VirtualLeaf} with explicit cell resolution, available in 2D, has been introduced in \cite{Merks2011}.
Although the cell representation is similar to vertex cell models, the dynamics is realized by minimizing an Hamiltonian using a Monte Carlo algorithm.
The model assumes rigid cell walls, which is appropriate for plant morphogenesis.

For many morphogenetic phenomena, which arise from a tight interaction between the biomolecular signalling and the tissue physics, an explicit computational representation of the cell shapes is required.
Here, we present a flexible software framework based on the IBCell model, which, as a novelty, permits to tightly couple biomolecular signalling models to a cell-resolved, physical tissue model.
The core components and the general approach of the model are described in the second section. In the third section, the software and the main functionalities are described in detail. Application examples are given in the fourth chapter to demonstrate the framework's capabilities.

\begin{figure}[t!]
\begin{centering}
\includegraphics[width=0.8\columnwidth]{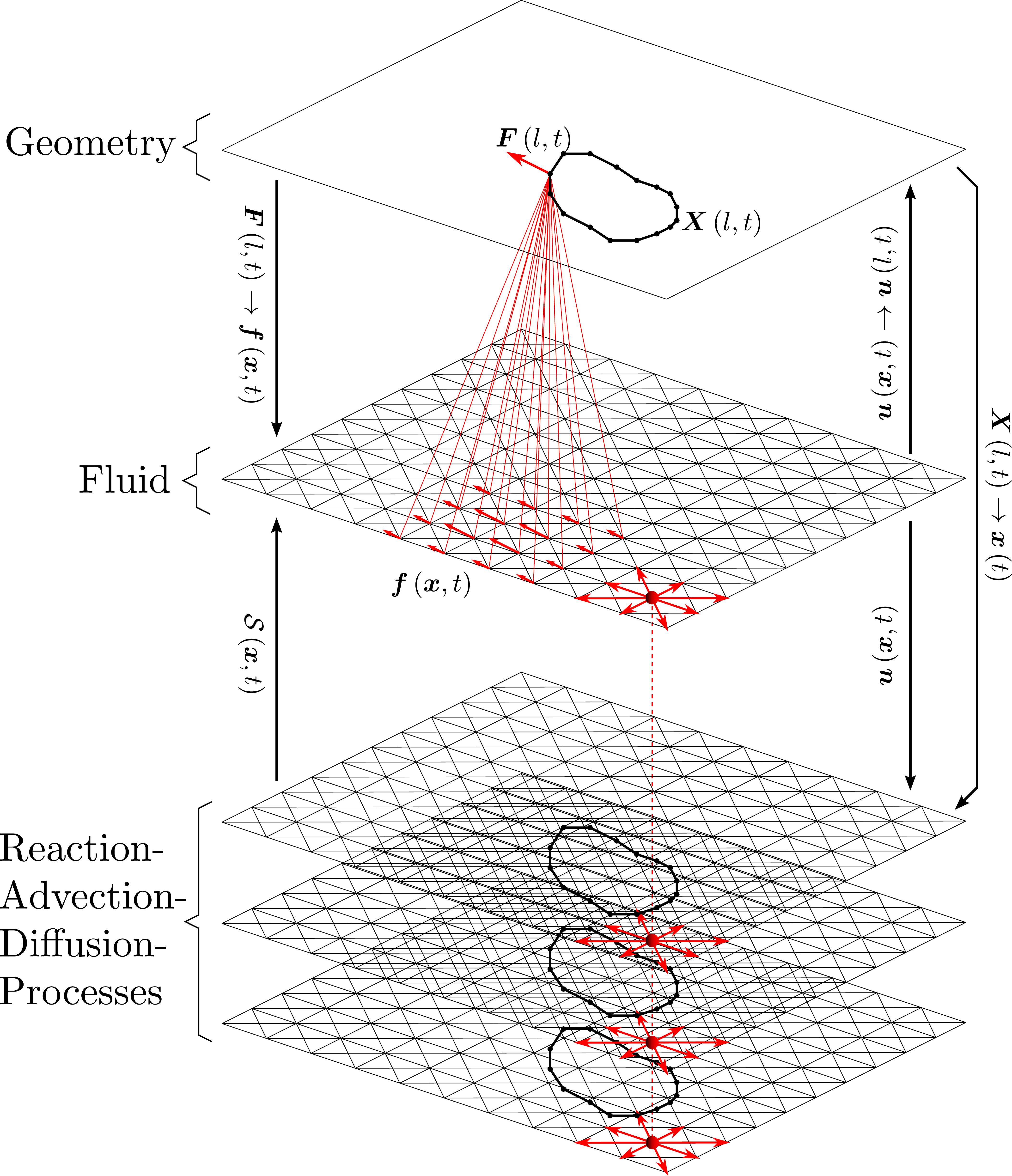}
\par\end{centering}
\caption{\textbf{Algorithm Overview.}
The algorithm consists of three coupled layers.
The geometry $\boldsymbol{X}\left(l,t\right)$ (top part, discussed in more detail in Figure \ref{fig:georepresentation}) is used to compute the forces {$\boldsymbol{F}\left(l,t\right)$} acting on each of the geometry nodes.
These forces, which do not necessarily coincide with a lattice point, are scattered to the fluid lattice (middle part) using the immersed boundary method kernel function, {$\boldsymbol{F}\left(l,t\right)\rightarrow \boldsymbol{f}\left(t\right)$}.
After advancing the fluid solver by one time step, the velocity is interpolated to the geometry node position using the same kernel function, $\boldsymbol{u}\left(\boldsymbol{x},t\right)\rightarrow \boldsymbol{U}\left(l,t\right)$.
The geometry nodes are moved according to their velocity $\boldsymbol{U}\left(l,t\right)$ and the iteration is restarted.
The velocity $\boldsymbol{u}\left(\boldsymbol{x},t\right)$ of the fluid lattice is also copied to the reaction-advection-diffusion solvers (PDE), together with the position $\boldsymbol{X}\left(l,t\right)\rightarrow \boldsymbol{x}\left(t\right)$ of the geometry.
The state of the reaction-advection-diffusion solvers, which are used to model signalling, may be used to compute mass sources $\mathcal{S}\left(\boldsymbol{x},t\right)$ for the fluid solver.}
\label{fig:algorithmoverview}
\end{figure}

\section{Approach}

Our approach permits the coupled simulation of tissue and signalling dynamics. To describe the tissue dynamics, the visco-elastic cell model needs to represent both the cellular structures and their elastic properties, as well as the viscous behaviour of the cytoplasm and of the extracellular space surrounding the cells.
The model therefore rests on three core parts: the representation of cells, the representation of the fluid and the fluid-structure interaction, and the coupling of the tissue part to the signalling model. To describe the interaction between the viscous fluid and the elastic structures, which are immersed in the fluid, we use the  immersed boundary (IB) method (\cite{Peskin2003}) as previously implemented in the visco-elastic cell model, also called IBcell model (\cite{Rejniak2004, Rejniak2007}).
To solve the viscous fluid behaviour, we use the Lattice Boltzmann method, which is an efficient mesoscopic numerical scheme, originally developed to solve fluid dynamics problems (\cite{Chen1998}).
The method has previously been successfully applied to reaction-diffusion equations such as Turing systems (\cite{PonceDawson1993}), as well as to coupled scalar fields such as temperature (\cite{Guo2002b}).
The method was for the first time combined with the immersed boundary method (\cite{Peskin2003}) by \cite{Feng2004}, and has later been used to study red blood cells in flow by \cite{Zhang2007}.
In the following, we provide an overview of the implemented methods; the implementation details are given in section \ref{sec:softwaredescription}.

%--------------------------------------------------------------------------------------------------------------------------------------------------------
\subsection{Cell Representation}
%--------------------------------------------------------------------------------------------------------------------------------------------------------

Cells are represented as massless, purely elastic structures, which are described by sets of geometry points forming polygons. The geometry points are connected via forces. In a first approximation, the elastic structures can be identified to represent the elastic cell membranes. However, more elastic structures can be added to the intra-/extracellular volume to mimic the visco-elastic properties of the cytoskeleton or the extracellular matrix. The user can implement biological mechanisms which operate on the cell representations. For example, a new junction to a neighboring cell might be created when the distance between two neighboring cell boundaries falls below a threshold distance. Similarly, a junction might be removed when overly stretched.

%--------------------------------------------------------------------------------------------------------------------------------------------------------
\subsection{Fluid and Fluid-Structure Interaction}
%--------------------------------------------------------------------------------------------------------------------------------------------------------
The visco-elastic cell model represents the content of cells (the cytoplasm) as well as the extracellular space (the interstitial fluid and the extracellular matrix) as a viscous, Newtonian fluid. The intra- and extracellular fluids interact with the elastic membrane, i.e. the fluids exert force on the membrane, and the membrane exerts force on the fluids. Furthermore, the velocity field of the fluid, which is induced by the forces, moves and deforms the elastic structures.
This interaction, well-known as fluid-structure interaction, lies at the heart of the tissue model. Forces (e.g. membrane tension or cell-cell forces) acting on these points are exerted on the fluid by distributing the force to the surrounding fluid.
Due to the local forcing, the fluid moves. 
At this step, the membrane point are advected passively by the fluid.
As a result the forces need to be re-evaluated on the points.
By repeating the forcing-advection steps, the interaction is realized iteratively.

As a result of this iterative process, the (elastic) structures are coupled to the (viscous) fluid. Depending on the parametrisation, this model allows to describe both elastic, or viscous, or visco-elastic material behaviour.
The upper part of Figure \ref{fig:algorithmoverview} illustrates the immersed boundary interaction.
The implemented IB kernel function has bounded support, i.e. each geometry point influences and is influenced only its immediate neighborhood.
Here, the dimension of the kernel function is four by four (cf. Figure \ref{fig:algorithmoverview}).
The fluid equations are solved using the Lattice Boltzmann method (\cite{Chen1998}), which is described in detail in the supplementary material (section \ref{sec:supplementarymaterial}).
The Reynolds number is typically $\lll 1$, hence the regime is described by Stokes flow\footnote{The Reynolds number reads $Re = \frac{UL}{\nu}$, with $U$ being a characteristic velocity,
$L$ a characteristic length scale, and $\nu$ the kinematic viscosity.
Assuming $L = 10^{-3} \, [m]$, $U = 10^{-8} \, [m/s]$ and $\nu = 10^{1} \dots 10^{2} \, [m^{2}/s]$, then
$Re = 10^{-13} \dots 10^{-12}$ can be estimated (\cite{Forgacs1998}).}.

%--------------------------------------------------------------------------------------------------------------------------------------------------------
\subsection{Signalling}
%--------------------------------------------------------------------------------------------------------------------------------------------------------

The signalling network is represented as a system of reaction-advection-diffusion processes.
The elastic membranes may act as no-flux boundaries for compounds which only exist in the extra- or intracellular volume, respectively.
The reaction-advection-diffusion solvers can be equipped with potentially coupled reaction terms in order to model signalling interactions of diffusing factors. Depending on the model, the signalling may impact the tissue dynamics.
This can be done, for instance, by making the mass source of the fluid dependent on the values of the reaction-advection-diffusion solvers such that the tissue expands locally (cf. Figure \ref{fig:algorithmoverview}).
Furthermore, the diffusing compounds can be individually configured to diffuse freely across the entire domain, or only inside or outside the cells (e.g. using no-flux boundary conditions for the cell membranes).

\begin{figure}[t!]
\begin{centering}
\includegraphics[width=0.8\columnwidth]{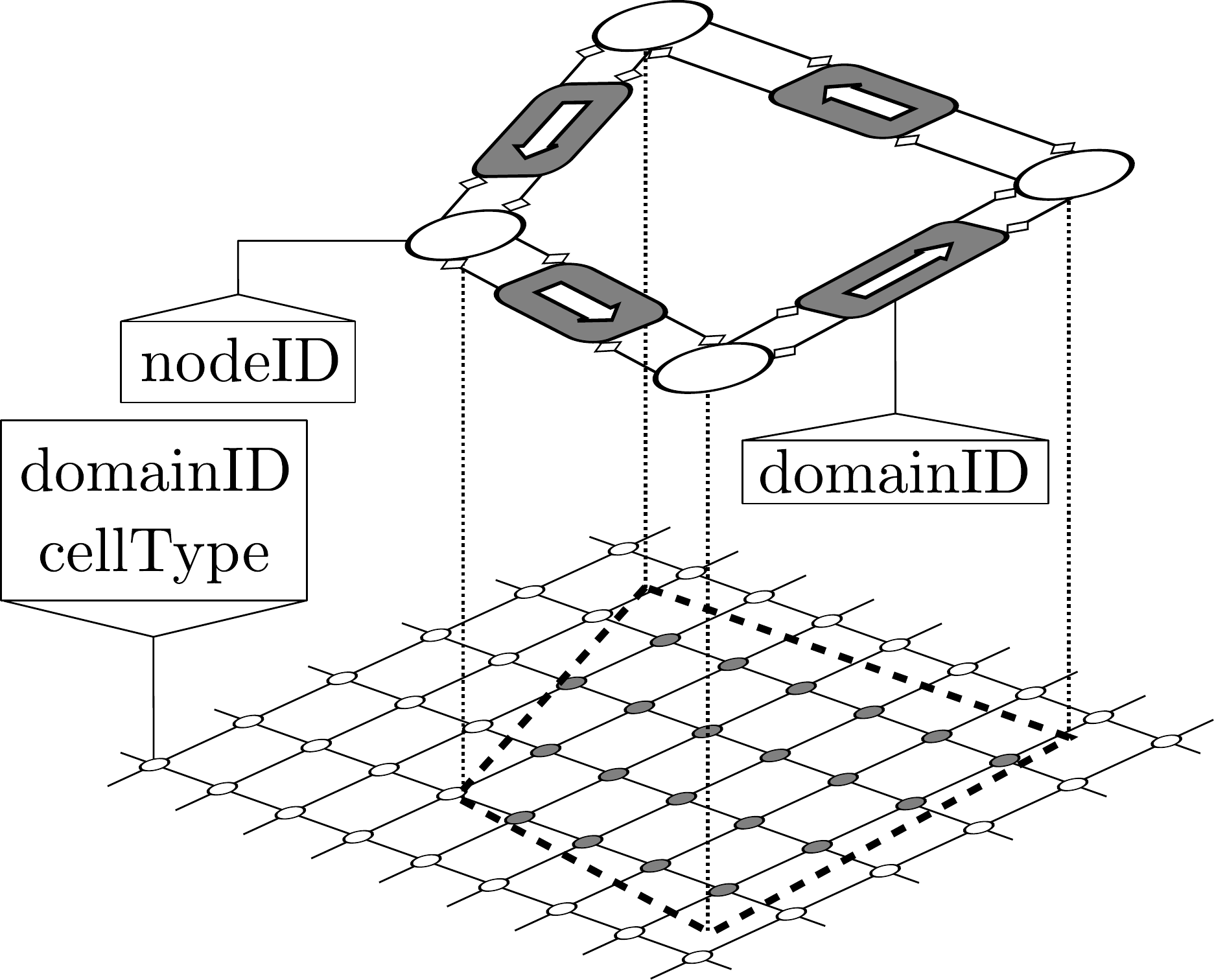}
\par\end{centering}
\caption{\textbf{Elements of the Geometry Representation.} 
The cells are closed polygons, consisting of geometry nodes (discs in the top part) and connections (shaded boxes in the top part) between each two geometry nodes.
Each connection stores two references to its preceding and successive geometry nodes, and \textit{vice versa} each geometry node stores two references to its preceding and successive connection (visualized by aggregation arrows in the top part).
Directionality of the polygon is counter clockwise by convention.
Each geometry node has a unique, immutable {\tt nodeID} attribute, which is allocated internally upon creation of a new geometry node.
Each connection features a {\tt domainID} attribute, which denotes the domain identifier of the domain on the left hand side.
The domain identifier on the right hand side is by definition zero, representing the extracellular space.
Using the {\tt domainID} of the connections, the {\tt domainID} of the lattice nodes is automatically set (lower part).
Additionally, each {\tt domainID} is associated with a {\tt cellType}.
The behaviour of the {\tt MassSolverXX}, {\tt BioSolverXX} and {\tt CDESolverXX} can be made dependent on the {\tt domainID} and/or {\tt cellType} attributes by the user.}
\label{fig:georepresentation}
\end{figure}

\begin{figure}[t!]
\begin{centering}
\includegraphics[width=0.8\columnwidth]{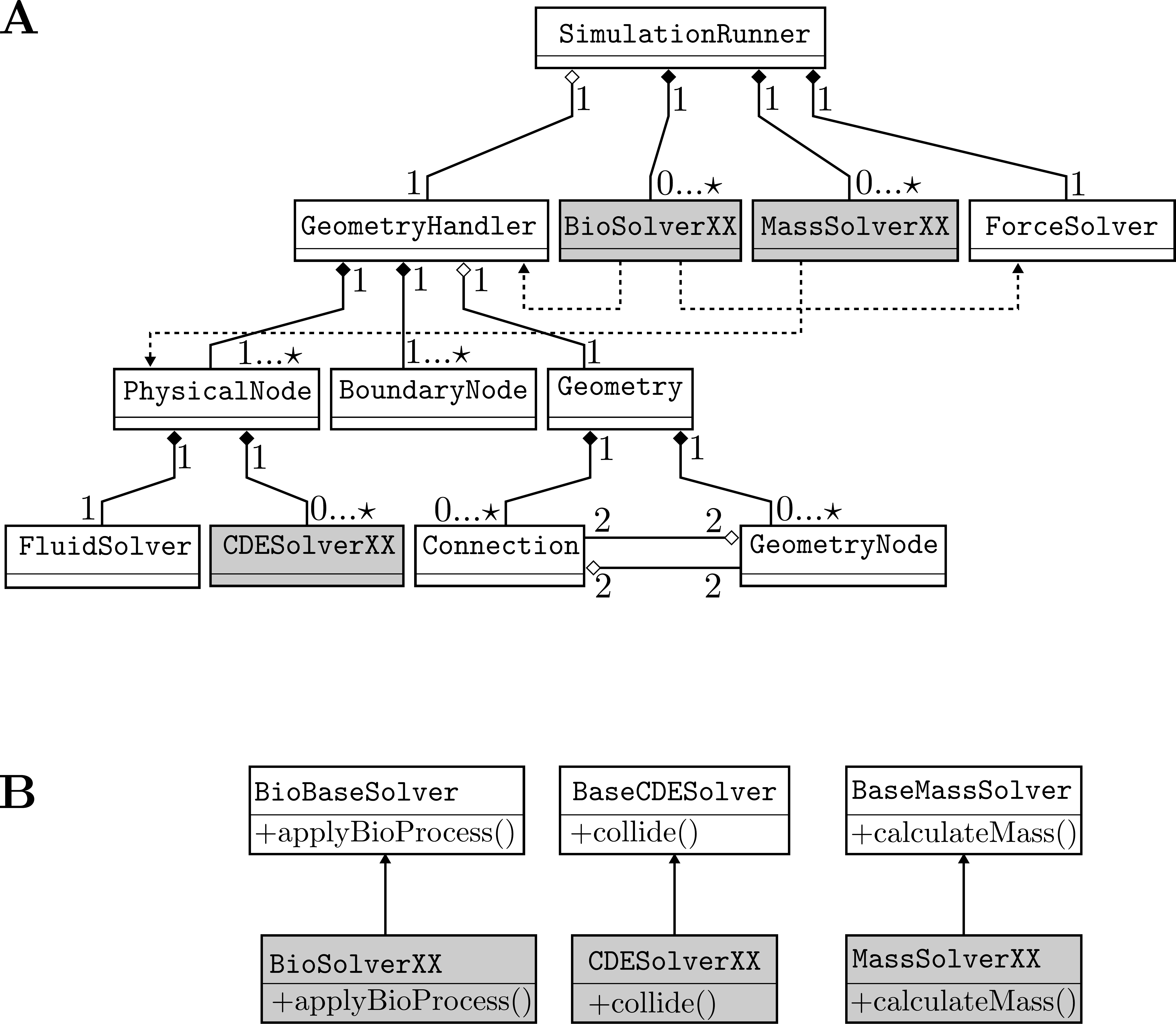}
\par\end{centering}
\caption{\textbf{Simplified UML diagram of important Classes.} 
The classes which have to be provided by the user are shaded.
{\tt XX} refers to an arbitrary solver name.
{\bf A} The {\tt SimulationRunner} controls the execution of the simulation.
The {\tt GeometryHandler}  has a collection of {\tt PhysicalNode}s, representing the lattice, a collection of {\tt BoundaryNode}s wich are woven into the lattice, and a {\tt Geometry} object.
The latter contains the cell's geometric information, namely the {\tt GeometryNode}s and the {\tt Connection}s.
The {\tt GeometryNode}s and the {\tt Connection}s each have two references of the preceding and successive elements, as also explained in Figure \ref{fig:georepresentation}.
{\tt BioSolverXX} obtains references from the {\tt GeometryHandler} and the {\tt ForceSolver} to alters states accordingly.
Similarly, the {\tt MassSolverXX} obtains a reference to the lattice and adds mass sources to the fluid.
{\bf B} To implement new custom routines, the user must inherit from provided base classes (from {\tt BioBaseSolver} for biologically motivated routines, from {\tt BaseCDESolver} for reaction-advection-diffusion processes, and from {\tt BaseMassSolver} for mass modifying routines)}
\label{fig:UML}
\end{figure}

%########################################################################################################################################################

\section{Software}\label{sec:softwaredescription}

%########################################################################################################################################################

%--------------------------------------------------------------------------------------------------------------------------------------------------------
\subsection{Cell Representation}
%--------------------------------------------------------------------------------------------------------------------------------------------------------

The cell geometries consist of two elements, the {\tt GeometryNode}s, which act as the IB points, and the {\tt Connection}s, connecting pairs of {\tt GeometryNode}s. A simplified cell is visualized in Figure \ref{fig:georepresentation}.
The {\tt Connection}s are attributed with a {\tt domainID} flag, which is an identifier for the surrounded domain (respecting the counter-clockwise directionality convention). The domain identifier on the other side (on the right hand side) is zero by convention, representing the interstitial space. The {\tt domainID} of the {\tt Connection}s are copied to the fluid and reaction-advection-diffusion solvers. Moreover, the {\tt domainID}'s are associated with a cell type flag, {\tt cellType}.
By applying custom differentiation rules, the {\tt cellType} of individual cells may be changed according to custom criteria; otherwise the all cells default to {\tt cellType}=1 (with {\tt cellType}=0 being the interstitial space, again).
In this way, the reaction terms and the mass sources may be made dependent on specific cells, or specific cell types.

%--------------------------------------------------------------------------------------------------------------------------------------------------------
\subsection{User-provided Solvers}
%--------------------------------------------------------------------------------------------------------------------------------------------------------

\begin{figure}[t!]
\begin{centering}
\includegraphics[width=0.7\columnwidth]{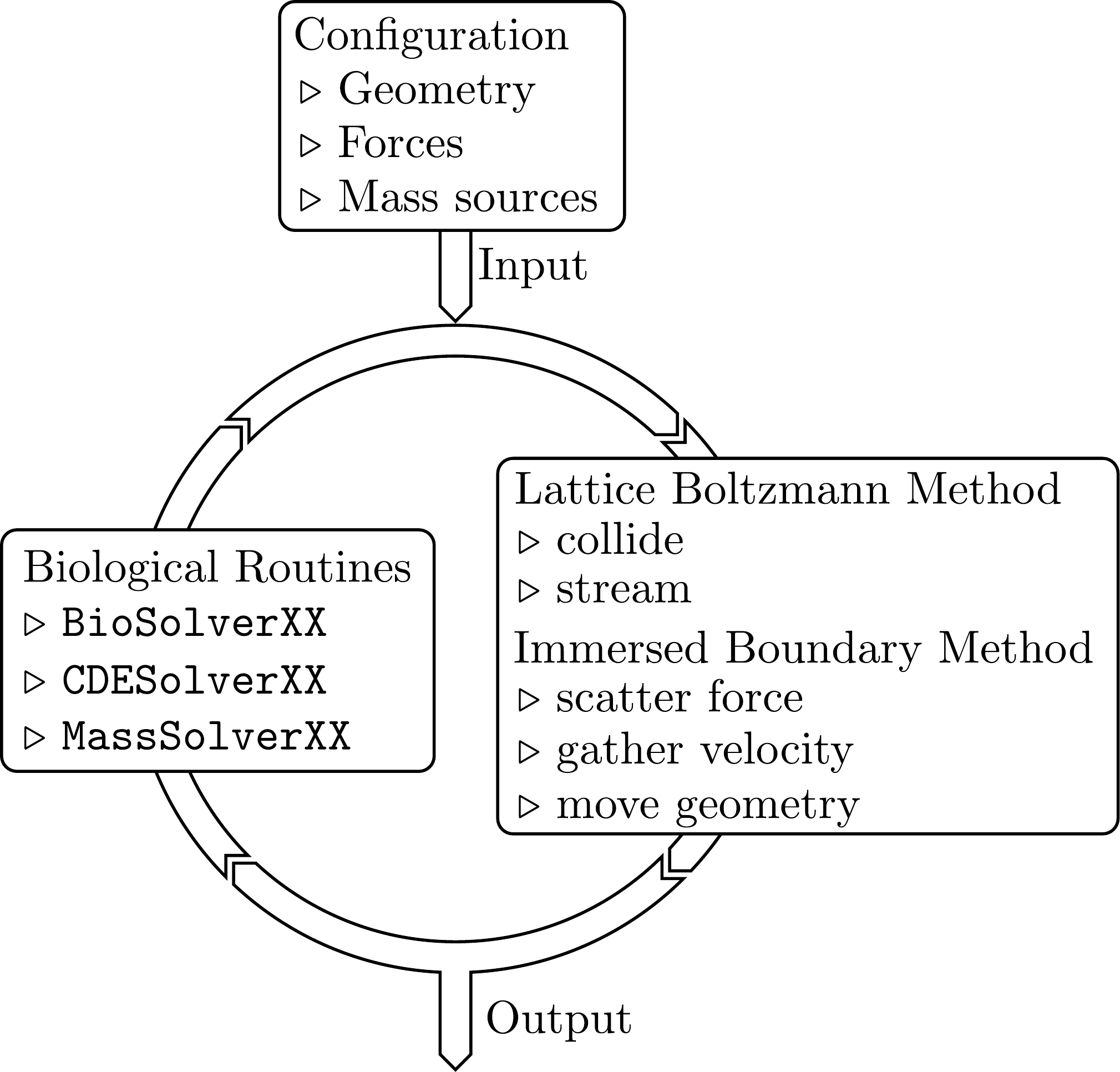}
\par\end{centering}
\caption{\textbf{Iterative Processing in the Solver} 
At initiation, the library loads the user-provided configuration files (containing global simulation parameters, initial geometry, initial forces).
During each iteration, the library's class {\tt SimulationRunner} (cf. Figure \ref{fig:UML}A) successively calls the physical routines (the Lattice Boltzmann method to solve the fluid and reaction-advection-diffusion processes, and the immersed boundary method to solve the fluid-structure interaction) and the biological routines (biologically motivated re-arrangement of the geometry, modifications of the forces, etc.). The current configuration and optionally the entire solver states can be saved at a chosen frequency.}
\label{fig:algorithmloop}
\end{figure}

The user can add the following routines: {\tt MassSolverXX}, {\tt CDESolverXX}, and {\tt BioSolverXX} ({\tt XX} being a name to be chosen).
The {\tt MassSolverXX} - as described above - adds or substracts mass from/to the fluid solver.
The {\tt CDESolverXX} is used to implement the reaction terms of the signalling models.
Finally, the {\tt BioSolverXX} can be used to execute biologically motivated operations on the geometry and the forces.
Such an operation might be cell division, which is discussed in more detail in Subsection \ref{subsubsec:celldivision}.
Figure \ref{fig:UML}A summarizes the most important classes and their interactions. The classes which are subject to customization are shaded.
In order to add a new customized routine (e.g. a mass modifying solver {\tt MassSolverXX}, a reaction-advection-diffusion solver {\tt CDESolverXX}, or a biologically motivated solver {\tt BioSolverXX}), the user needs to inherit from their respective virtual base classes (cf. Figure \ref{fig:UML}B).
Figure \ref{fig:algorithmloop} visualizes the routines, which are called iteratively by the {\tt SimulationRunner} (cf. Figure \ref{fig:UML}A).

%--------------------------------------------------------------------------------------------------------------------------------------------------------
\subsection{Input and Output}
%--------------------------------------------------------------------------------------------------------------------------------------------------------

The communication to the user is achieved via the loading and dumping of configuration files.
A general configuration file contains the global simulation parameters, such as the simulation time, the domain size, the fluid viscosity, and the diffusion coefficients for the reaction-advection-diffusion solvers.
The geometry points and the corresponding geometrical connections are stored in a geometry file. 
A third file contains the forces, including forces between a pair of geometry points, freely defined forces or spatially anchored points.
The fluid and reaction-advection-diffusion solver states may be written either to .txt files or in .vtk format and can be post-processed with third-party software (e.g. Matlab or ParaView).
Optionally, the solver states can be saved in a loadable format to resume the simulation.

%--------------------------------------------------------------------------------------------------------------------------------------------------------
\subsection{Physical Processes}
%--------------------------------------------------------------------------------------------------------------------------------------------------------

\subsubsection{Viscous and Elastic Behaviour}
The viscous behaviour is implemented using a representation of an incompressible fluid (solved using the Lattice Boltzmann method, cf. supplementary material), which converges to the Navier-Stokes equation in the hydrodynamic limit.
The fluid is solved on a regular Cartesian and Eulerian grid.
The membranes are represented by sets of points, which are connected to form closed polygons.
A variety of forces may act on the membrane nodes, such as e.g. membrane tensions (cf. Subsection \ref{subsec:forces}).
The interaction between the fluid and the elastic structures is formulated using the Immersed Boundary method (cf. supplementary material).
The membrane points move according to the local fluid velocity field in a Lagrangian manner.

\subsubsection{Reaction-Diffusion of Biochemical Compounds}
The biochemical signalling can be described by sets of coupled reaction-diffusion partial differential equations.
Similar to the fluid equations, these equations are solved on a regular Cartesian and Eulerian grid (solved using the Lattice Boltzmann method, cf. supplementary material).
The concentrations of the compounds can be accessed by other solvers, for example to make other processes such as cell division dependent on signalling factors.
The cell boundaries can be chosen to be either invisible to the diffusing compounds, or to be no-flux boundaries.
To account for advection, the fluid velocity field is directly transferred from the fluid solver since the fluid and the reaction-diffusion lattices coincide spatially.
The coupling of the solvers is visualized in Figure \ref{fig:algorithmoverview}.

\subsubsection{Forces}\label{subsec:forces}
Forces are an integral part of the simulation environment. A force is always connected to a membrane point. Any type of conservative force (which can be derived from a potential) can easily be implemented. Currently, the following types of forces are implemented:
\begin{itemize}
 \item spring force between two geometrical nodes
 \item spring force between a geometrical node and a spatial anchor point
 \item free force acting on a geometrical node
 \item horizontally or vertically sliding force (thus enforcing only the y or x coordinate, respectively)
 \item constant force between two geometrical nodes
\end{itemize}
Application examples include constant forces between two geometrical nodes that can be used to model constant membrane tension, which leads to the minimization of a cells perimeter (discussed in Subsection \ref{subsubsec:membranetension}).
Moreover, a geometrical point can dynamically explore its local neighbourhood and establish a force to another geometrical point \textit{from another cell}, thus mimicking cell-cell junctions (discussed in Subsection \ref{subsubsec:celljunctions}).

%--------------------------------------------------------------------------------------------------------------------------------------------------------
\subsection{Biological Processes}\label{subsec:biologicalprocesses}
%--------------------------------------------------------------------------------------------------------------------------------------------------------

The biological solvers (BioSolver) accommodate the functionalities that are related to biological processes.
These processes may be mostly related to modifications of the forces and the geometry.
The BioSolver has full access to the compound concentrations.
Furthermore, it is aware of the cells, whose geometries are stored individually.
This enables the BioSolver to compute cell areas and averaged or integrated compound concentrations.
Since all cells are individually tagged, cell behaviour can be made dependent on cell identity.
Additionally to the cell identity, cells also carry a cell type tag, which can be changed depending on run-time conditions.
This latter functionality can be used to model cell differentiation. 

Consider a cell division event as an example. Here, a division plane has to be chosen. The choice of its position and direction is subject to the user's model: the cell division plane might be set perpendicular to the cell's axis of strongest elongation. Next, the cell has to be divided, which requires the removal of the corresponding geometrical connections, and the insertion of new geometrical nodes and connections to close the divided cells.

Note that the concentration fields of the compounds, as well as the velocity- and pressure fields of the fluid solver are not directly altered in the biological module.

\subsubsection{Control of Cell Area}\label{subsubsec:cellarea}

Depending on the biological model of the user, the cell area has to be controlled.
By assuming that a cell might change its spatial extent in the third dimension, the area might shrink or expand as a response to forces exerted by its neighbouring cells, which can effectively be modelled as an 'area elasticity'.
In the limiting case, the cell resists external forces, maintains its area and only reacts with changes of the hydrostatic pressure.
In general, to control the area of cells, the reference area for each cell needs to be adapted.
The reference area acts as a set point for a simple proportional controller, i.e. the local mass source $\mathcal{S}_{k}$ in the cell $k$ is proportional to the area difference between the current cell area $A_{k}\left(t\right)$, and the set point area $A_{k}^{0}$:
\begin{equation}
\mathcal{S}_{k} = \alpha \left( A_{k}^{0} - A_{k}\left(t\right)\right)
\end{equation}
where $\alpha$ is a proportional constant.
More advanced control methods, such as e.g. proportional-integral control methods, can be realized easily.

This approach of controlling the cell area can also be used to let cells grow or shrink in a controlled way, i.e. a cell differentiating into an hypertrophic cell type may grow in volume. Implementing this process would be as simple as setting the new target area as set point area. The area controller will bring the cell close to its new area.

\subsubsection{Membrane Tension}\label{subsubsec:membranetension}
The definition of forces acting between pairs of membrane points allows for simulating the cell's membrane tension.
By default, a constant contracting force $\boldsymbol{F_{i}}$ with magnitude $\varphi^{\mbox{m}}$ is applied to every pair of neighbouring membrane points.
Hence the resulting force on membrane point $i$ is composed of a force pointing to its preceding membrane point $i-1$, and a force pointing to its successive membrane point $i+1$:
\begin{equation}\label{eq:membraneforce}
\boldsymbol{F_{i}^{m}} = \varphi^{m} \left( \frac{\boldsymbol{x}_{i-1} - \boldsymbol{x}_{i}}{ \left| \boldsymbol{x}_{i-1} - \boldsymbol{x}_{i} \right|} +
						\frac{\boldsymbol{x}_{i+1} - \boldsymbol{x}_{i}}{ \left| \boldsymbol{x}_{i+1} - \boldsymbol{x}_{i} \right|} \right)
\end{equation}

This approach can be interpreted as an actively remodelled membrane: when stretched, new membrane is synthesized in order to not increase the membrane tension on longer time scales (hours).
On the other hand, excessive membrane is degraded to abide the membrane tension.
Therefore, the membrane tension minimizes the cell's perimeter.
Since the intracellular fluid (and thus the cell area) is conserved in the absence of neighbouring cells and active mechanisms (c.f. Subsection \ref{subsubsec:cellarea}), the cell assumes a circular shape.
On short time scales (seconds), the passive (non-remodelled) elastic membranes can be modelled by using Hookean spring potentials.
The membrane tension will then be proportional to deviation from the resting membrane perimeter.
In both cases, the membrane is flexible (i.e. has no bending stiffness); if bending stiffness should be required by the user, this can be easily realised in a custom {\tt BioSolver}.

The implementation of membrane tension needs to consider the geometry remeshing. Whenever a new membrane point is inserted, it needs to get connected to its neighbours instantly, because the cell will be overly stretched in the absence of membrane tensions. A membrane point's forces need to be removed upon its removal. Algorithmically, this is realized by removing and reconstructing all membrane forces at every time step. At this point, the magnitude of the membrane tension can be made dependent on signalling factors.

{\tt BioSolverMembraneTension} is an example of a class managing the membrane tensions with immediate remodeling, and\\
\noindent{\tt BioSolverHookeanMembraneTension} implements simple Hookean springs.

\subsubsection{Cell Junctions}\label{subsubsec:celljunctions}
A cell can create cell junctions to neighbouring cells.
In the simplest case, each membrane point $i$ uses the function \texttt{getGeometryNodesWithinRadiusWithAvoidance-} \texttt{Closest} to get the closest membrane point $j$ of another cell, which is within a predefined cut-off radius $l^{max}$, or zero if there is no such membrane point. Once a candidate membrane point fulfils the criteria, a new Hookean force $\boldsymbol{F_{i}}$ with a spring constant $k^{\mbox{j}}$ and resting length $l_{0}$ is created:
\begin{equation}\label{eq:celljunctionforce}
\boldsymbol{F_{i}^{cj}} =
\begin{cases}
k^{\mbox{j}} \frac{\boldsymbol{x}_{j}-\boldsymbol{x_{i}}}{\left| \boldsymbol{x}_{j}-\boldsymbol{x_{i}} \right|} \left(\left| \boldsymbol{x}_{j}-\boldsymbol{x_{i}} \right| - l_{0} \right) &\text{if} \,
\left| \boldsymbol{x}_{j}-\boldsymbol{x_{i}} \right| < l^{max}
\\
0 & \text{else}
\end{cases}
\end{equation}

The cell junction forces are regularly (potentially not at every time step) deleted and renewed, where the frequency of cell junction renewal might reflect the cell junction synthesis rate.

The function \texttt{getGeometryNodesWithinRadius-}\\ \texttt{WithAvoidance} returns all membrane points of another cell, which are within a predefined cut-off radius; the returned list might be empty. This opens up the possibility to introduce randomness by choosing the membrane point randomly from the candidate list.
The probability to create a junction might depend on the junction length: the shorter, the higher the probability to form a new junction.
Also the removal of membrane points might be randomized, and the probability made dependent on the junction length, i.e. overly stretched junctions are removed with higher probability. Even the membrane point whose junctions shall be updated might be chosen randomly. Again, the number of updated membrane nodes per time reflects the cell's limited cell junction synthesis activity.

The membrane points are internally stored in an fast neighbor list data structure, which is well suited for spatial range queries.
{\tt BioSolverCellJunction} is an example of a class responsible for cell junctions.

\subsubsection{Cell Division}\label{subsubsec:celldivision}
The cell division functionality requires several steps.
First, criteria will have to be defined which cells shall be divided.
Criteria might be maximal cell area, maximal spatial expansion, or biochemical signals.
Once a cell committed for division, the cell division plane will have to be chosen.
Again, how to chose the plane is subject to biological modelling.
A frequently used rule is to use a plane defined by a random direction vector and the center of mass of the cell.
However, different rules can be readily implemented, such as random directions drawn from non-uniform probability distributions (which, in turn, can be controlled e.g. by signalling factor gradients) or division planes perpendicular to the longest axis (\cite{Minc2011}).
In a next step, the two membrane segments are determined which intersect with the division plane; this is implemented in {\tt getTwoConnectionsRandomDirection} or {\tt getTwoConnectionsLongestAxis}.
These two membrane segments are subsequently removed, and two new membrane segments across the cell are introduced, leading to a cut through the mother cell.
Finally, a new domain identity number has to be given to one of the daughter cells; the other daughter cell inherits the domain identity number from the mother cell. The new domain identity number is set to the largest domain identity number plus one, and it is automatically copied to the physical grid. Both daughter cells by default inherit the cell type flag from the mother cell, which is also automatically copied to the physical grid.

The basic cell division functionality is implemented in the class {\tt BioSolverCellDivision}.

\subsubsection{Differentiation}
Differentiation changes the cell type flag of the cells according to user-defined, biologically motivated rules.
These rules might be based on the cell area, or on a signalling factor concentration, possibly integrated over the cell area.
Once being committed for differentiation, the cell changes its cell type flag according to the rule.
The new cell type flag will be automatically copied to the physical grid.
The cell type flag can be used to make signalling dynamics, but also other biologically motivated processes dependent on the cell type.

The association between the domain identifier flags and the cell type flags is stored in the {\tt cellTypeTrackerMap\_}, which is a member of the {\tt GeometryHandler}.
This makes sure that all {\tt BioSolverXX} classes have easy access to this information.
A basic implementation of the differentiation control can be found in {\tt BioSolverDifferentiation}.

%--------------------------------------------------------------------------------------------------------------------------------------------------------
\subsection{Accuracy and Performance}
%--------------------------------------------------------------------------------------------------------------------------------------------------------

The Lattice Boltzmann schemes are second order accurate, and the explicit immersed boundary method is first order accurate in space and time.
The internal data structure uses a fast neighbor list (cell list) implementation to optimize for range queries (e.g. searching for other cells in the local neighborhood), which exhibits a search complexity of $\mathcal{O}\left(N\right)$, with $N$ being the number of membrane points to represent the cells.
Many iterative computations (LB and IB routines such as particle streaming and collision, gathering of velocity and scattering of force) are parallelized using the shared memory paradigm. However, a few computational steps cannot be parallelized.
This is typically the case when write-operations occur on shared data structures, such as the data structures storing the geometry nodes and the force structs (e.g. in {\tt ForceSolver::deleteForceType()} and {\tt GeometryHandler::computeAreas()}).
Moreover, the geometry remeshing (refining and coarsening) functions as well as the data I/O are not parellelized, but are assumed to occur much less frequently than the actual fluid and reaction-advection-diffusion solvers.
Therefore, since the fraction of sequential code is not negligible, the software should best be run on fast multi-core processors.

%--------------------------------------------------------------------------------------------------------------------------------------------------------
\subsection{Tools, Dependencies and Documentation}
%--------------------------------------------------------------------------------------------------------------------------------------------------------

A compiler with C++0x support (such as GCC 4.7 or higher) is required.
The software depends on Boost (\url{http://www.boost.org}; 1.54.0 or higher), OpenMP, CMake (\url{http://www.cmake.org}) and vtk (\url{http://www.vtk.org/}; 5.8 or higher).
The source code is extensively documented using Doxygen (\url{http://www.stack.nl/~dimitri/doxygen}).
Git (\url{http://git-scm.com}) is used for version control.
The software has only been tested on linux operating systems.

%--------------------------------------------------------------------------------------------------------------------------------------------------------
\subsection{Availability}
%--------------------------------------------------------------------------------------------------------------------------------------------------------

The documentation and source code are available on\\
{\url{http://tanakas.bitbucket.org/lbibcell/index.html}}

%########################################################################################################################################################
\section{Application Examples}
%########################################################################################################################################################

\begin{figure}[t!]
\begin{centering}
\includegraphics[width=0.8\columnwidth]{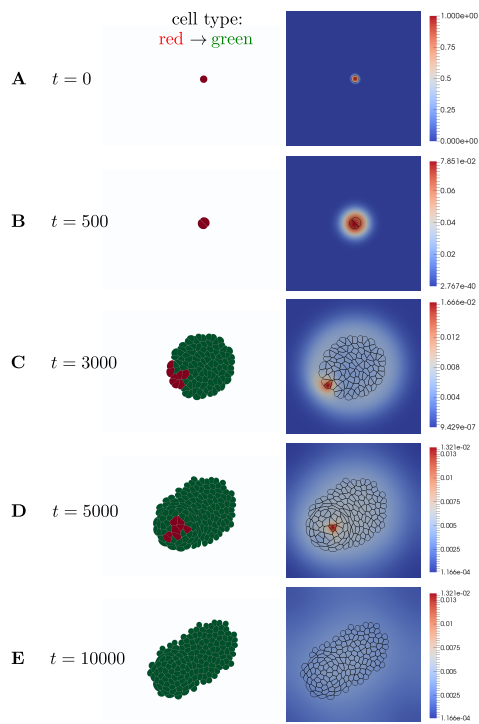}
\par\end{centering}
\caption{\textbf{Cell Division, Differentiation and Signalling.}
{\bf A} The initial configuration consists of a single, circular cell of type red.
The red cell type proliferates at a high rate.
The initial cell is tagged and expresses a signalling molecule $\mathcal{I}$ which inhibits differentiation.
{\bf B} The first cell division occurs. The division axis is chosen randomly.
The daughter cell inherits the cell type from the mother cell, but only the mother cell keeps expressing the signalling molecule $\mathcal{I}$.
{\bf C} The signalling level (the spatially integrated concentration of the signalling molecule) drops in cells far away from the initial cell and differentiation into the green cell type occurs.
The green cell type does not grow intrinsically, and only divides if overly stretched by external forces.
{\bf D} The highly proliferating red cells are trapped in the forming tissue due to the randomly chosen cell division axis.
At $t=5000$, the expression of the differentiation inhibiting molecule $\mathcal{I}$ is switched off, which leads to the differentiation of the remaining red cells.
{\bf E} In the absence of high proliferation, the cells rearrange to maximize the perimeter/area ratio. Characteristic penta- and hexagonal cell shapes emerge (cf. Supp. \ref{Farhadifar}). Cells close to the boundary try to take a circular shape.}
\label{fig:example_panel}
\end{figure}

%--------------------------------------------------------------------------------------------------------------------------------------------------------
\subsection{Cell Division, Differentiation and Signalling}
%--------------------------------------------------------------------------------------------------------------------------------------------------------

To demonstrate the capabilities of the software we first consider a tissue model with cell-type specific cell division and signalling-dependent differentiation (Figure \ref{fig:example_panel}). In the beginning, a circular cell with radius $R=10$ is placed in the middle of a quadratic $400$ by $400$ domain (Figure \ref{fig:example_panel}A). Iso-pressure boundary conditions are set at the border of the domain.
The initial cell is of red cell type, which is proliferating at a high rate.
When considering a single layer epithelium,
mass uptake, which is needed for modelling cell growth and finally proliferation, is assumed to occur from the apical cavity through the apical membrane.
Additionally, the initial cell secretes a signalling factor $\mathcal{I}$ which inhibits differentiation of the red cell type into the green cell type.
Once the cell area doubled, the cell is divided in a random direction (cf. Figure \ref{fig:example_panel}B).
The daughter cells inherit the cell type, but only the mother cell continues to express the signalling molecule $\mathcal{I}$.
All cells of red type integrate the concentration of $\mathcal{I}$ over their area.
For low signalling levels, the red cell type differentiates into the green cell type.
The green cell type does not grow and only divides if external forces stretch the cell.
In Figure \ref{fig:example_panel}C, the daughter cell's signalling level dropped after cell division, and differentiation occurred.
After several rounds of cell division, a tissue starts to form (cf. Figure \ref{fig:example_panel}D). The cells close to the secreting initial cell remain protected from differentiation, whereas more distant cells differentiate irreversibly.
Due to the randomly chosen cell division axis, it might happen that the proliferating red cells get trapped (cf. Figure \ref{fig:example_panel}E).
The expression of $\mathcal{I}$ is switched off at time t=5000, thus leading to complete differentiation shortly after (cf. Figure \ref{fig:example_panel}F).
After proliferation stopped, the cells slowly rearrange because 
cell-cell junctions are broken if overly stretched, and new junctions are formed (according to Eq. (\ref{eq:celljunctionforce}) ).
At the boundary of the tissue, the cells try to reach a spherical shape, while in the middle mainly characteristic penta- and hexagonal shapes emerge (cf. Figure \ref{fig:example_panel}F and Supp. \ref{Farhadifar}).

\begin{figure}[t!]
\begin{centering}
\includegraphics[width=\columnwidth]{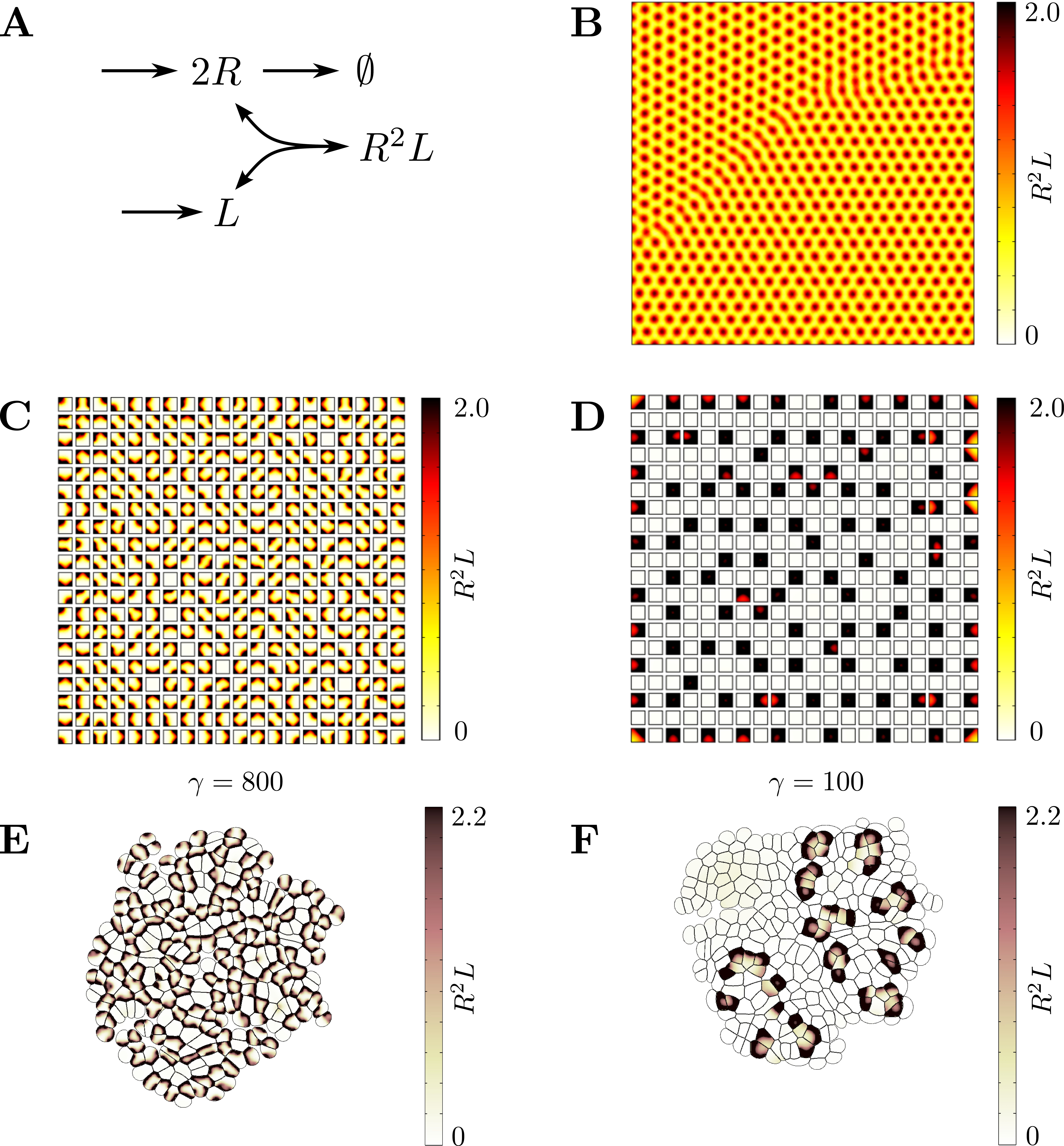}
\par\end{centering}
\caption{\textbf{Turing Patterning on Growing Cellular Domains} {\bf A} Turing instability can be achieved by Schnakenberg-type reactions, involving a slowly diffusing compound $R$, here interpreted as a receptor, and a fast diffusing compound $L$, here interpreted as a freely diffusing ligand.
One ligand molecule binds to two receptors, leading to the complex $R^{2}L$. The complex can be interpreted as a biological signal.
{\bf B} The model is solved on a continuous square lattice (using $d=1$, $\gamma=800$, $a=0.1$, $b=0.9$), resulting in the classical regular spot-pattern. The biological signal $R^{2}L$ is shown.
{\bf C} The same system as in B is solved on an idealized static cellular domain, i.e. the diffusion of the receptor $R$ is restricted to a cell. The emerging biological signal $R^{2}L$ is now distributed irregularly.
{\bf D} The same system as in C, but with $\gamma=100$, is solved on a idealized static cellular domain.  Fewer cells show significant levels of signal $R^{2}L$ and no regular pattern can be found (salt-and-pepper pattern).
{\bf E} The same system as in C is solved on a growing cellular domain.
The proliferation rate of a cell is set proportional to its signal $R^{2}L$.
The resulting pattern features regularity on a larger scale, but the local patterning significantly differs from the behaviour on continuous (B) and static cellular (C) domains.
{\bf F} The same system as in D is solved on growing cellular domain.
The proliferation rate of a cell is set proportional to the local intensity of the signal $R^{2}L$.
Clusters of active cells with high levels of $R^{2}L$ emerge.}
\label{fig:cellbased_turing}
\end{figure}

%--------------------------------------------------------------------------------------------------------------------------------------------------------
\subsection{Turing Patterning on Growing Cellular Domains}
%--------------------------------------------------------------------------------------------------------------------------------------------------------

To demonstrate the importance to investigate morphogenic signalling hypotheses on dynamically growing domains with cellular resolution, we solved a reaction-diffusion system, featuring the well-known diffusion-driven Turing instability (\cite{Turing1952a}), on a proliferating tissue. Figure \ref{fig:cellbased_turing}A illustrates the interaction between a ligand $\mathrm{L}$ and its receptor $\mathrm{R}$.
Here, we assume that one ligand dimer molecule $\mathrm{L}$ binds to two receptors $\mathrm{R}$, forming the complex $\mathrm{R}^{2}\mathrm{L}$ which induces upregulation of the receptor on the membrane (e.g. \cite{Bellusci1997}).
Unbound receptor is turned over at a linear rate.
The ligand can diffuse freely across the tissue and the entire domain, whereas the diffusion of the receptor is limited to a single cell's apical surface and is much slower. The dynamics can be formulated as a system of non-dimensional partial differential equations:
\begin{align}
\partial_{t} R &= \Delta R + \gamma \left( a - R  + R^{2} L \right)\\
\partial_{t} L &= d\Delta L + \gamma \left( b - R^{2} L \right)
\end{align}
where $\gamma$ is a reactivity constant, $a$ and $b$ production constants, and $d$ the relative diffusion coefficient of ligand and receptor.
We note that the equations correspond to the classical Schnakenberg-type Turing mechanism (\cite{Gierer1972,Schnakenberg1979}).
It has previously been shown that such a receptor-ligand interaction can explain symmetry breaking in various morphogenetic systems (\cite{Menshykau2012,Celliere2012,Badugu2012,Menshykau2013,Tanaka2013,MenshykauDBlancPUnalE}).

Depending on the type of domain we observe very different patterns.
On a continuous domain we obtain the well known regular spot pattern (Figure \ref{fig:cellbased_turing}B).
On an idealized static cellular domain an overall regular pattern with irregular internal structure (Figure \ref{fig:cellbased_turing}C) can be observed.
Decreasing the simulation parameter $\gamma$, which inversely controls the distance between the spots, leads to even more unexpected patterns:
for $\gamma=100$, the local regularity is completely lost (Figure \ref{fig:cellbased_turing}D).
Finally, on a dynamically growing cellular domain, where the local proliferation rate was set proportional to the $\mathrm{R}^{2}\mathrm{L}$ signal, we obtain irregular patterns (Figure \ref{fig:cellbased_turing}E).
For a lower value $\gamma=100$, clusters of cells with high $\mathrm{R}^{2}\mathrm{L}$ signalling levels emerge (Figure \ref{fig:cellbased_turing}F).
In conclusion, even relatively simple signalling mechanisms can lead to significantly different results, depending on how the tissue is represented.

%--------------------------------------------------------------------------------------------------------------------------------------------------------
%--------------------------------------------------------------------------------------------------------------------------------------------------------
\section{Discussion}
%--------------------------------------------------------------------------------------------------------------------------------------------------------
%--------------------------------------------------------------------------------------------------------------------------------------------------------

We developed an extendible and open-source cell-based simulation environment, which is tailored to study morphogenetic problems.
The novel framework permits the coupled simulation of  a physically motivated visco-elastic cell model with regulatory signalling models.
Processes such as viscous dissipation, elasticity, advection, diffusion, local reactions, local mass sources and sinks, cell division and cell differentiation are implemented.
By applying our framework to Turing signalling systems we show that the signalling systems may behave very differently on dynamic tissues than on simple continuous tissue representations.
We therefore advocate to test continuous morphogenetic signalling models on dynamically growing cellular domains.

The presented framework permits to study a variety of mechano-regulatory mechanisms.
By making the cell division orientation dependent on signalling cues, the effect on the macroscopic tissue geometry may be studied.
Cell migration can be modelled by introducing gradient-dependent forces on specific cell types. Cell sorting may be achieved by specifying multiple cell types with differential cell-cell junction strengths.
The framework is specifically designed to study the mutual effects of signalling and biophysical cell properties.

The visco-elastic cell model represents cell shapes at very high resolution and is thus, unlike the vertex model, not restricted to densely packed tissues.
Furthermore, hydrodynamic interaction, membrane tension and hydrostatic pressure are integral components of the model.
The fact that a velocity field is available on the entire domain is a critical advantage to account for advection of the signalling components, thus allowing for a spatial description of intracellular concentrations.
The model is, however, not easily extendable to the third dimension.
Since a meshing of the surface will be required, the algorithmic and computational complexity are expected to be significant and subject to future work.
The presented framework is, however, ideal to study intrinsically two-dimensional morphogenetic problems, such as apical surface dynamics of epithelia as studied previously also by \cite{Farhadifar2007} and
\cite{Ishihara2012a} in 2D.

\section*{Acknowledgement}

\paragraph{Funding}
The authors acknowledge funding from the SNF Sinergia grant "Developmental engineering of endochondral ossification from mesenchymal stem cells" and the SNF SystemsX RTD NeurostemX.

\clearpage
\bibliographystyle{plain}
\bibliography{MyCollection}

\clearpage
%########################################################################################################################################################
\section{Supplementary Material}\label{sec:supplementarymaterial}
%########################################################################################################################################################

\subsection{Lattice Boltzmann Method}\label{subsec:lbm}

For the fluid, the standard Lattice Boltzmann scheme (with the single-relaxation-time Bhatnagar-Gross-Krook collision operator) is used (\cite{Chen1998}).
It has been shown, that the incompressible Navier-Stokes equations can be recovered in the hydrodynamic limit (\cite{Chen1998}).
The Boltzmann equation is discretized on a
D2Q9\footnote {the nine-velocity lattice in two dimensions is defined as
$\boldsymbol{v}_{0} = \left[0,0\right]$,
$\boldsymbol{v}_{i}= \left[  \cos\left(\pi \left(i-1\right)/2\right) , \sin\left( \pi \left(i-1\right)/2  \right) \right]$
for $i=\{1,2,3,4\}$, and
$\boldsymbol{v}_{i}= \sqrt{2} \left[ \cos\left(\pi \left(i-1\right)/2\right) + \pi/4 , \sin\left( \pi \left(i-1\right)/2 + \pi/4 \right) \right]$
for $i=\{5,6,7,8\}$ }
lattice and reads:
\begin{multline}\label{eq:bgkequation}
 f_{i} \left( \boldsymbol{x}+\boldsymbol{v}_{i},t+1\right) - f_{i}\left(\boldsymbol{x},t\right) = \\
 - \frac{1}{\tau} \left( f_{i}\left(\boldsymbol{x},t\right) - f_{i}^{eq}\left(\boldsymbol{x},t\right) \right)
 + \Delta t \frac{w_{i} \rho}{c_{s}^{2}} \left( \boldsymbol{v}_{i} \cdot \boldsymbol{f} \right)
\end{multline}
where $f_{i}$ denotes the particle distribution function in the direction $i$,
and can be interpreted as the probability density of finding particles with velocity $\boldsymbol{v}_{i}$ at time $t$ at position $\boldsymbol{x}$.
The relaxation time $\tau$ is related to
the kinematic viscosity $\nu = c_{s}^{2}\left(\tau - \frac{1}{2}\right)$.
For isothermal flows, the speed of sound $c_{s}$ is defined as $c_{s}=c/\sqrt{3}$
using the lattice speed $c=\Delta x/\Delta t$.
The lattice spacing and the time step are chosen as $\Delta x=\Delta t =1$ to guarantee consistency with the lattice.
The last term of Equation (\ref{eq:bgkequation}) represents the external body force (\cite{He1997}).

Equation (\ref{eq:bgkequation}) implies a two step algorithm: firstly, the distribution functions $f_{i}$ perform a free flight to the next lattice point (left hand side).
Secondly, on each lattice point, the incoming distribution functions collide and relax towards a local equilibrium distribution $f_{i}^{eq}$,
which is controlled by the relaxation time $\tau$ (right hand side).
The equilibrium distribution is taken as:

\begin{equation}\label{eq:standardequilibrium}
 f_{i}^{eq} = \omega_{i} \rho \left[ 1 + \frac{ \boldsymbol{v}_{i} \cdot \boldsymbol{u} } {c_{s}^{2}} 
		    + \frac{\left(\boldsymbol{v}_{i} \boldsymbol{v}_{i}-v^{2}\right) : \boldsymbol{u} \boldsymbol{u}} {2 c_{s}^{2}} \right]
\end{equation}
with the fluid velocity $\boldsymbol{u}$ and the fluid density $\rho$. The operator $:$ denotes the dyadic tensor scalar product.
For the D2Q9 lattice, the population weights $\omega_{i}$ can be found as $\omega_{0}=4/9$, $\omega_{1-4}=1/9$, and $\omega_{5-8}=1/36$.

Finally, the macroscopic quantities (density $\rho$ and momentum density $\rho \boldsymbol{u}$) can be computed using the zeroth and first order moments:
\begin{equation}
\rho = \sum_{i=0}^{8} f_{i} \hspace{2cm} \rho\boldsymbol{u} = \sum_{i=0}^{8} f_{i}\boldsymbol{v}_{i}
\end{equation}
The fluid pressure $p$ is related to the mass density $\rho$ as $p=\rho c_{s}^{2}$.

For solving the reaction-advection-diffusion equation of a compound $\phi$, a multi-distribution function approach is chosen (\cite{BARTOLONI1993,Guo2002b}), i.e.
an additional Lattice Boltzmann solver is coupled to the fluid solver.
We implemented different schemes (D2Q4,D2Q5), which exhibit slightly different stability and accuracy (\cite{Li2001}).
For the D2Q5
\footnote{
the five-velocity lattice in two dimensions is defined as
$\boldsymbol{v}_{0} = \left[0,0\right]$ and
$\boldsymbol{v}_{i}= \left[  \cos\left(\pi \left(i-1\right)/2\right) , \sin\left( \pi \left(i-1\right)/2  \right) \right]$
for $i=\{1,2,3,4\}$
}
scheme, we follow \cite{Huber2008} and \cite{Parmigiani2009} and write for the Lattice Boltzmann equation:
\begin{equation}
g_{i} \left( \boldsymbol{x}+\boldsymbol{v}_{i},t+1\right) - g_{i}\left(\boldsymbol{x},t\right) =
 - \frac{1}{\tau_{D}} \left( g_{i}\left(\boldsymbol{x},t\right) - g_{i}^{eq}\left(\boldsymbol{x},t\right) \right)
\end{equation}
The equilibrium distribution functions is taken as:
\begin{equation}
g_{i}^{eq} = \phi w_{i} \left[ 1 + \frac{1}{c_{s}^{2}} \left( \boldsymbol{v}_{i} \cdot \boldsymbol{u} \right) \right]
\end{equation}
Instead of the local first moment, the velocity field $\boldsymbol{u}$ from the fluid is transferred.
The weights $w_{i}$ are chosen as $w_{0}=1/3$ and $w_{1-4}=1/6$.
The relaxation parameter $\tau_{D}$ is related to the diffusion coefficient as
$D = c_{s}^{2} \left( \tau_{D} - \frac{1}{2}\right)$,
and the local compound density $\phi$ reads:
\begin{equation}
\phi = \sum_{i=0}^{4}g_{i}
\end{equation}

All variables are expressed in Lattice Boltzmann units $\delta x, \delta t$ and have to be converted to physical units.

\subsubsection{No-flux Boundary Condition for Reaction-Advection-Diffusion Equations}

The missing incoming distribution functions are approximated by equilibrium distributions.
The momentum is spatially first order interpolated
between the fluid in direction of the missing population, and the known zero momentum at the wall. The density is spatially first order extrapolated
from the fluid.

\subsubsection{Pressure Boundary Condition for the Fluid Equations}
Within the computational domain, lattice points can act as internal pressure boundaries, i.e. points with prescribed fluid pressure, 
which may be needed to implement mass sinks in case of growing cells/tissues.
All distribution functions are rescaled such that the prescribed pressure is obtained.
The velocity field is not affected, since the rescaling factor $\gamma$ cancels:

\begin{equation}
 \boldsymbol{u}\left(\boldsymbol{x},t\right) = \frac{\sum_{i}\gamma \boldsymbol{v}_{i} f_{i} \left(\boldsymbol{x},t\right)}
    {\sum_{i} \gamma f_{i}\left(\boldsymbol{x},t\right)}
\end{equation}

\subsection{Immersed Boundary Method}\label{subsec:ibm}

The PDE's are prefentially solved on a Eulerian grid, whereas cells are naturally represented in a Lagrangian manner.
The two frames can be bridged using the immersed boundary method, which was introduced in \cite{Peskin1977}.
This technique is appropriate for complex and moving boundaries and fluid-structure interactions.
The material equations are solved on a Eulerian grid, whereas the boundary equations are expressed in a Lagrangian way.
An introduction can be found in \cite{Mittal2005},
and detailed discussions in \cite{Peskin2003}.

The curvlinear (boundary) coordinates $\left(q,r,s\right)$ are attached to a material point. At time $t$, the coordinates
in the Eulerian framework read $\boldsymbol{X}\left(q,r,s,t\right)$.
The material derivative $\frac{D \boldsymbol{u}}{Dt} \left(\boldsymbol{x},t\right) = \frac{\partial^{2} \boldsymbol{X}}{\partial t^{2}} \left(q,r,s,t\right)$ is the acceleration of
whatever material point is at position $\boldsymbol{x}$ at time $t$.
The conversion from Lagrangian variables (e.g. mass density $M\left( q,r,s,t\right)$ or force density $\boldsymbol{F}\left(q,r,s,t\right)$) to 
their Eulerian counterparts is executed by integrating over the material coordinates:
\begin{eqnarray}\label{eq:conversion1}
 \rho \left(\boldsymbol{x},t\right) =
  \int M\left(q,r,s,t\right)\delta\left( \boldsymbol{x}-\boldsymbol{X}\left(q,r,s,t\right) \right)  dq dr ds\\
 \boldsymbol{f} \left(\boldsymbol{x},t\right) = \int \boldsymbol{F}\left(q,r,s,t\right) \delta\left( \boldsymbol{x}-\boldsymbol{X}\left(q,r,s,t\right) \right)  dq dr ds \label{eq:forcedistribution}
\end{eqnarray}
where $\delta\left(\cdot\right)$ denotes the delta Dirac function.
For the numerical implementation, the Delta Dirac function is approximated in such a way that it covers multiple grid points, on which the
conversions in Equation (\ref{eq:conversion1}) and (\ref{eq:conversion2}) are evaluated.
The equation of motion of the Lagrangian particles reads:
\begin{multline}\label{eq:conversion2}
 \frac{\partial \boldsymbol{X}}{\partial t}\left(q,r,s,t\right) = \\ \boldsymbol{u}\left(\boldsymbol{X}\left(q,r,s,t\right),t\right)
 = \int \boldsymbol{u}\left(\boldsymbol{x},t\right) \delta \left( \boldsymbol{x}-\boldsymbol{X}\left(q,r,s,t\right) \right) d\boldsymbol{x}
\end{multline}

Using the Hodge decomposition\footnote{any vector field $\boldsymbol{u}$ can be written
as $\rho \frac{D\boldsymbol{u}}{Dt}-\boldsymbol{f} = -\nabla p + \boldsymbol{w}$, where $\nabla \cdot \boldsymbol{w} =0$},
the material is described by

\begin{eqnarray}\label{eq:navierstokes2}
 \rho \left[ \frac{\partial \boldsymbol{u}}{\partial t} + \left(\boldsymbol{u} \cdot \nabla\right) \boldsymbol{u} \right] &=& -\nabla p + \mu\Delta \boldsymbol{u} + \boldsymbol{f}\\
 \nabla \cdot \boldsymbol{u} &=& 0
\end{eqnarray}

To couple the Immersed Boundary method to the Lattice Boltzmann method, the LB forcing term $\boldsymbol{f}$ has to be obtained from the Lagrangian force $\boldsymbol{F}$ following a similar way as \cite{Feng2004}.
The Lagrangian force $\boldsymbol{F}_{i}$ of the geometry point $i$ is composed of membrane forces (cf. Equation (\ref{eq:membraneforce})) and cell junction forces (cf. Equation (\ref{eq:membraneforce})):
\begin{equation}
\boldsymbol{F}_{i} = \boldsymbol{F}_{i}^{m} + \boldsymbol{F}_{i}^{cj}
\end{equation}
$\boldsymbol{F}_{i}$ is distributed to the local fluid neighborhood according to Equation (\ref{eq:forcedistribution}), and the Eulerian force $\boldsymbol{f}$ on the lattice point $\boldsymbol{x}$ is then used in the LB collision term in Equation (\ref{eq:bgkequation}).

The discretized delta Dirac function for one spatial dimension is defined as:
\begin{equation}\label{eq:1Dkernel}
\delta^{1D}\left( \boldsymbol{r}\right) =
\begin{cases}
   \frac{1}{4} \left(1+\cos\left(\frac{\pi \|\boldsymbol{r} \|}{2}\right)\right) & \text{if } \|\boldsymbol{r} \| \leq 2 \\
   0       & \text{if } \|\boldsymbol{r} \|>2
  \end{cases}
\end{equation}
and the two-dimensional discretized delta Dirac function is a multiplication of Equation (\ref{eq:1Dkernel}):
\begin{equation}
\delta\left(\boldsymbol{r}\right) = \delta^{1D}\left( \boldsymbol{r}\right) \delta^{1D}\left( \boldsymbol{r}\right)
\end{equation}
This implies that the Lagrangian force $\boldsymbol{F}_{i}$ of a membrane point $i$ is distributed to the $4\times 4=16$ nearest fluid lattice points.
By discretizing the membrane into mebrane points $i$, the discretized delta Dirac functions overlap.

%--------------------------------------------------------------------------------------------------------------------------------------------------------
\subsection{Validations}
%--------------------------------------------------------------------------------------------------------------------------------------------------------

To validate the implementation of the fluid solver, the reaction-advection-diffusion solver and the immersed boundary method,
validation tests using problems with known analytical solutions are executed.

The dimensionless problem is as follows: a circular cell (diameter 1) is placed in the middle of a square domain of size 10 by 10.
The end simulation time is $T=1$.
Inside the cell, a diffusing species is defined, where the cell membrane acts as a no-flux boundary condition.
No production and degradation of the species occurs.
The initial condition of the species is set to be uniformly 1.
The diffusion coefficient as well as the kinematic viscosity of the fluid are set to 1.
No membrane forces are applied; iso-pressure outflow boundary conditions are applied to the domain boundaries.
Please see \textit{movie\_S3.avi} for an impression of how the concentration decreases as the area increases.

\begin{figure}[t!]
\begin{centering}
\includegraphics[width=\columnwidth]{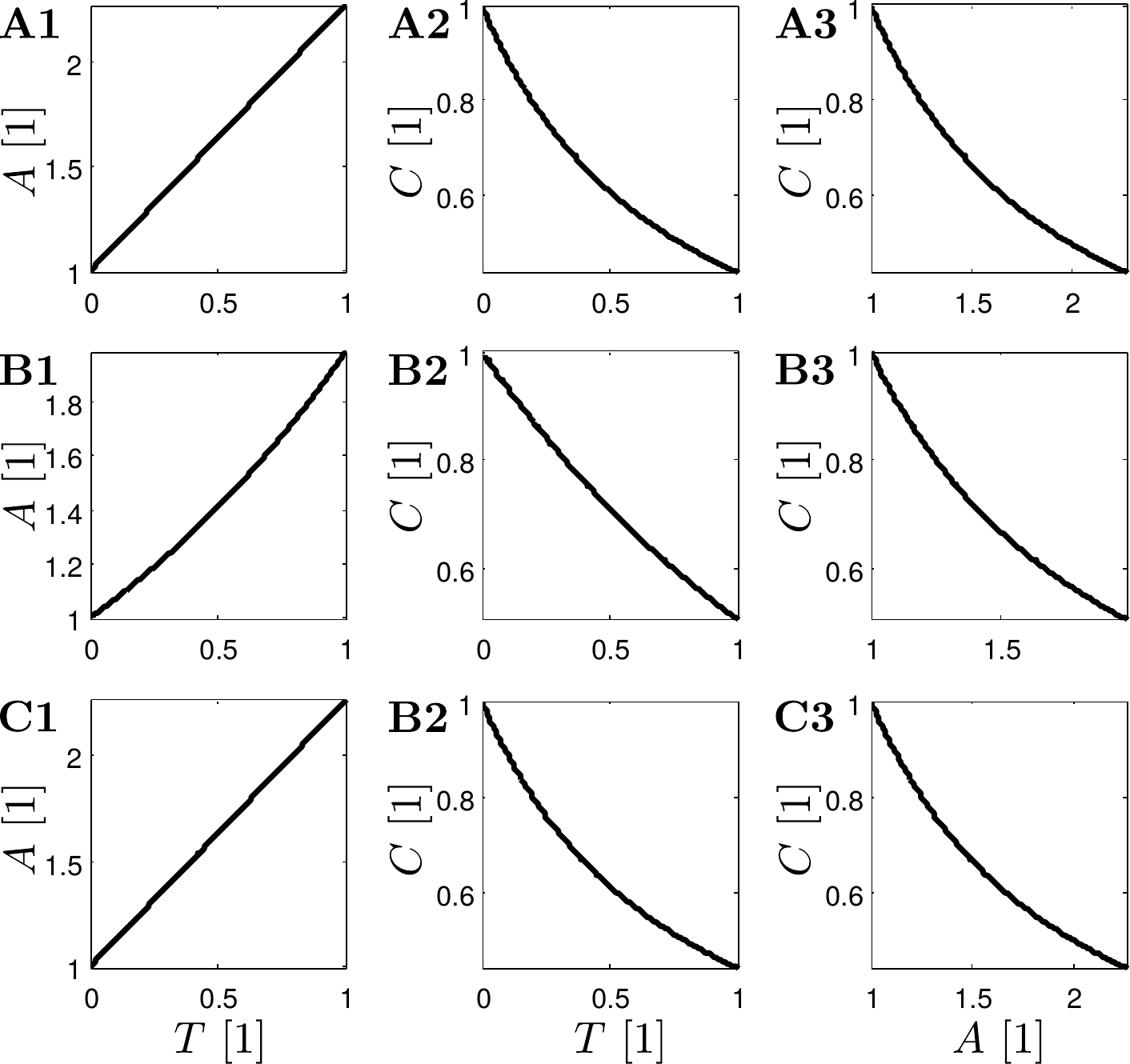}
\par\end{centering}
\caption{\textbf{Validation Cases: Evolution of Area and Concentration}
$T$ is the non-dimensional normalized time, $A$ the non-dimensional normalized area, $C$ the non-dimensionalized normalized concentration.
The column {\bf X1} shows area $A$ vs. time $T$; column {\bf X2} shows concentration $C$ vs. time $T$; and column {\bf X3} shows concentration $C$ vs. area $A$.
{\bf A} Case I: constant point-like mass source;
{\bf B} Case II: constant uniform mass source;
{\bf C} Case III: uniform mass source proportional to the concentration $C$.
}
\label{fig:figure_S1}
\end{figure}

Three different scenarios are considered:
\begin{itemize}
\item Case I: a constant point mass source is placed in the middle of the cell (dimensionless strength is 1); see Fig. \ref{fig:figure_S1}A
\item Case II: a constant uniform mass source inside the cell (dimensionless strength is $\log(2)/A_{0}$, where $A_{0}=0.5^{2}\pi$ is the initial cell area; this leads to a doubling of the cell area in time $T$), see Fig. \ref{fig:figure_S1}B
\item Case III: a uniform mass source inside the cell, where the strength is scaled linear proportional to the species concentration ($=1\cdot C(t)$, where $C(t)$ is the concentration of the species; this leads to a linear mass/area increase in time),  see Fig. \ref{fig:figure_S1}C
\end{itemize}

For all three cases, the circular cell starts to grow, and the species is diluted, i.e. the species concentration drops as the area increased.
Ideally, the integrated species concentration is conserved.
The relative errors of the simulations w.r.t. the analytical solutions are computed for both the area and concentration evolution.

The simulations are executed at different levels of space and time discretization.
As a measure, we take the diameter resolution of the circular cell, which is set to $\{10,20,40,80\}$.
As a result, the discretized values ($LB$ denotes Lattice Boltzmann units, and $ND$ denotes non-dimensional) for all cases
can be found in Table \ref{table:lbdiscretization}.
\begin{table}[h]\scriptsize
\begin{tabular}{|l|l|l|l|l|}
\hline
circle diameter & 10 & 20 & 40 & 80 \\
\hline
domain size & 100x100 & 200x200 & 400x400 & 800x800 \\
\hline
$\tau$ fluid & 1 & 1 & 1 & 1 \\
\hline
$\nu^{LB}$ fluid & 1/6 & 1/6 & 1/6 & 1/6 \\
\hline
$\tau$ diff & 1 & 1 & 1 & 1\\
\hline
$\nu^{LB}$ diff & 1/6 & 1/6 & 1/6 & 1/6 \\
\hline
$\delta_{x}$ & 0.1 & 0.05 & 0.025 & 0.0125 \\
\hline
$T^{LB}$ & 1e4 & 4e4 & 1.6e5 & 6.4e5 \\
\hline
$\delta_{t}$ & 1e-4 & 25e-6 & 625e-8 & 15625e-10 \\
\hline
$\mathcal{S}^{LB} = \mathcal{S}^{ND}\cdot \frac{\delta_{t}}{\delta_{x}^{2}}$ & 1e-2 & 1e-2& 1e-2 & 1e-2 \\
\hline
\end{tabular}
\caption{Lattice-Boltzmann discretization of Cases I-III.
ND and LB denote non-dimensional and Lattice Boltzmann units, respectively}
\label{table:lbdiscretization}
\end{table}

\begin{figure}[t!]
\begin{centering}
\includegraphics[width=\columnwidth]{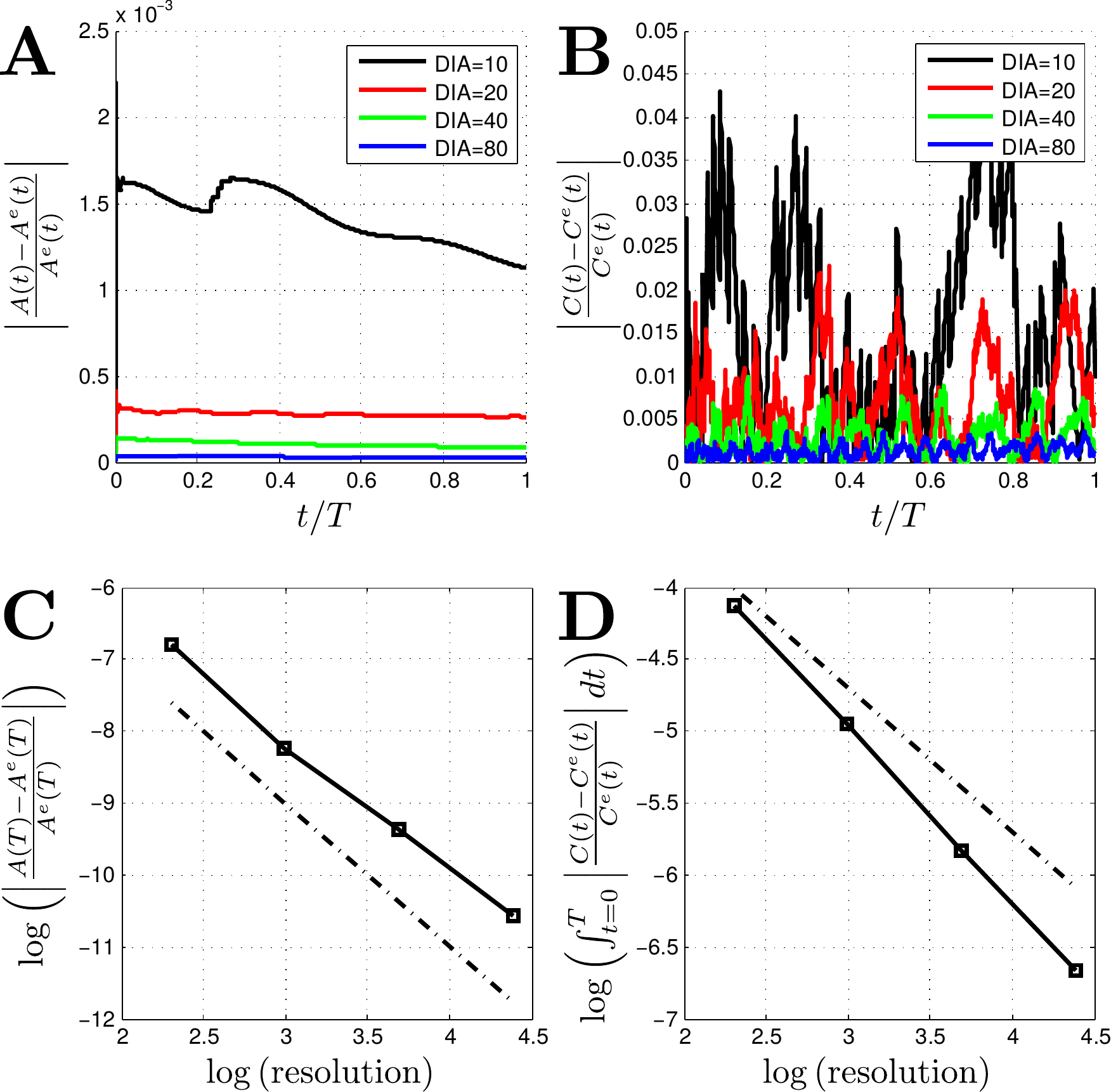}
\par\end{centering}
\caption{\textbf{Case I Validation: Point Source.}
An initially circular cell is growing through addition of mass.
The mass is added by using a point source in the middle of the domain.
Four different spatial resolutions have been computed, corresponding to a resolution of $\mbox{DIA}=\{10,20,40,80\}$ lattice points for the diameter of the initial circular cell.
{\bf A} The time evolution of the relative error of the cell area for 4 different lattice resolutions.
{\bf B} The time evolution of the relative error of the spatially integrated concentration (mass conservation) for 4 different lattice resolutions.
{\bf C} Convergence plot of the relative error of the cell area as a function of the lattice resolution.
The fitted slope is -1.80. The dashed line denotes a slope of -2.
{\bf D} Convergence plot of the relative error of the integrated concentration (mass conservation) as a function of the lattice resolution.
The fitted slope is -1.22. The dashed line denotes a slope of -1.
}
\label{fig:figure_S2}
\end{figure}

The validation results for Case I are given in Figure \ref{fig:figure_S2}.
Figure \ref{fig:figure_S2}{\bf A} shows the relative error temporal evolution of the cell area for the different resolutions,
and \ref{fig:figure_S2}{\bf B} the relative error temporal evolution of the integrated concentration.
The concentration error is much more noisy because it is only evaluated on lattice points.
When the cell grows, the membrane sweeps across the lattice, and whenever a new lattice site enters the intracellular domain, the 
integrated concentration increases abruptly.
To compute the convergence w.r.t. to spatial and temporal resolution, the relative errors at $T^{ND}=1$ are evaluated for the area, 
but the temporal mean is taken for the concentration convergence analysis.
Figure \ref{fig:figure_S2}{\bf C} shows the convergence analysis for the area.
For the finer grids, the convergence order is approximately 1.
Figure \ref{fig:figure_S2}{\bf D} depicts the convergence analysis for the integrated concentration,
where the fitted order of convergence is $\approx 1.2$.

\begin{figure}[t!]
\begin{centering}
\includegraphics[width=\columnwidth]{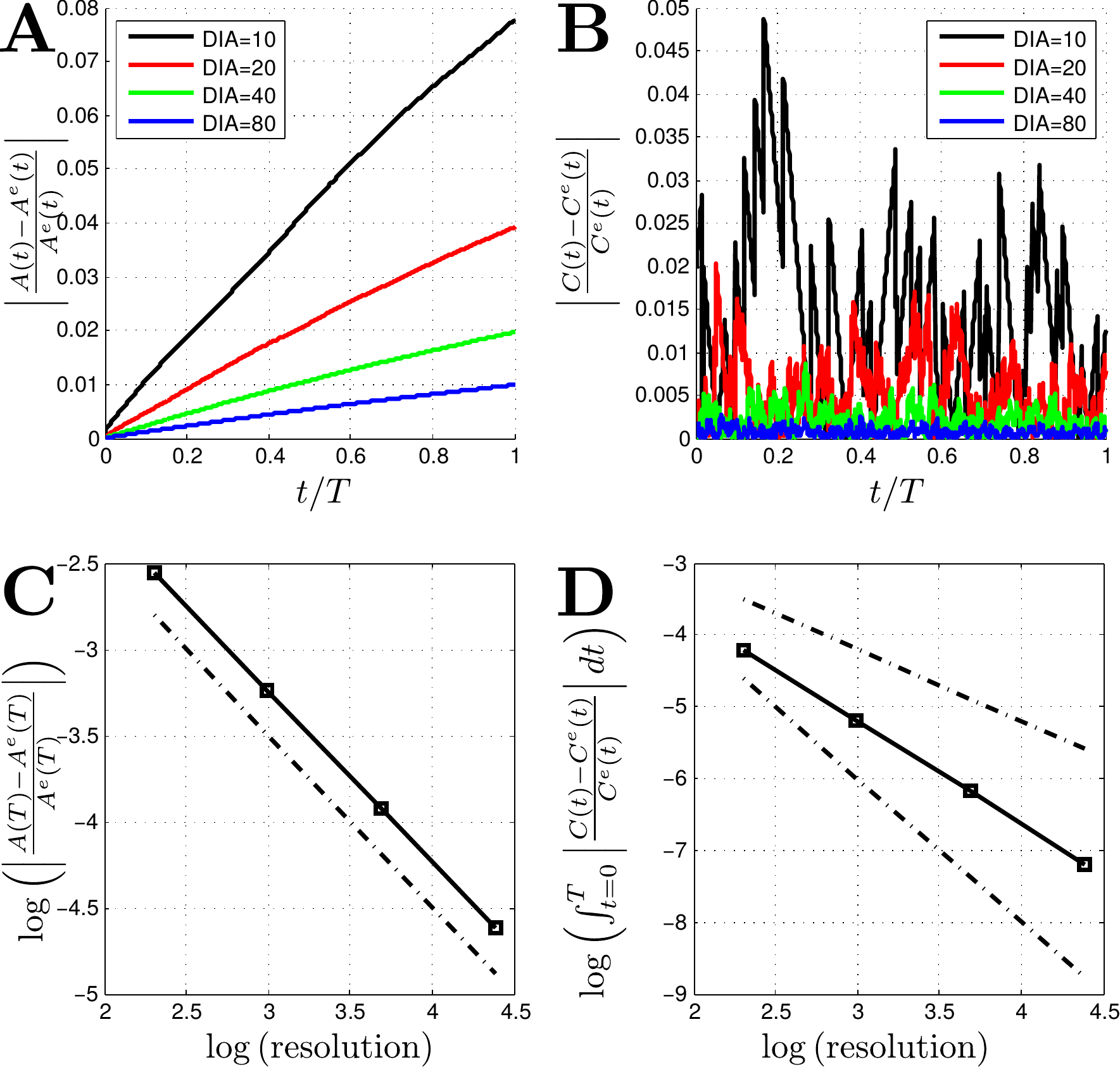}
\par\end{centering}
\caption{\textbf{Case II Validation: Uniformly Distributed Source.}
An initially circular cell is growing through addition of mass.
The mass is added by applying a constant uniform mass source on the entire cellular domain.
Four different spatial resolutions have been computed, corresponding to a resolution of $\mbox{DIA}=\{10,20,40,80\}$ lattice points for the diameter of the initial circular cell.
{\bf A} The time evolution of the relative error of the cell area for 4 different lattice resolutions.
{\bf B} The time evolution of the relative error of the spatially integrated concentration (mass conservation) for 4 different lattice resolutions.
{\bf C} Convergence plot of the relative error of the cell area as a function of the lattice resolution.
The fitted slope is -0.99. The dashed line denotes a slope of -1.
{\bf D} Convergence plot of the relative error of the integrated concentration (mass conservation) as a function of the lattice resolution.
The fitted slope is -1.43. The dashed lines denote slopes of -1 and -2, respectively.
}
\label{fig:figure_S3}
\end{figure}

The validation results for Case II are given in Figure \ref{fig:figure_S3}.
Figure \ref{fig:figure_S3}{\bf A} shows the relative error temporal evolution of the cell area for the different resolutions,
and \ref{fig:figure_S3}{\bf B} the relative error temporal evolution of the integrated concentration.
The concentration error is much more noisy because the it is only evaluated on lattice points.
When the cell grows, the membrane sweeps across the lattice, and whenever a new lattice site enters the intracellular domain, the 
integrated concentration increases abruptly.
To compute the convergence w.r.t. to spatial and temporal resolution, the relative errors at $T^{ND}=1$ are evaluated for the area, 
but the temporal mean is taken for the concentration convergence analysis.
Figure \ref{fig:figure_S3}{\bf C} shows the convergence analysis for the area.
For the finer grids, the convergence order is approximately 1.
Figure \ref{fig:figure_S3}{\bf D} depicts the convergence analysis for the integrated concentration,
where the fitted order of convergence is $\approx 1.4$.

\begin{figure}[t!]
\begin{centering}
\includegraphics[width=\columnwidth]{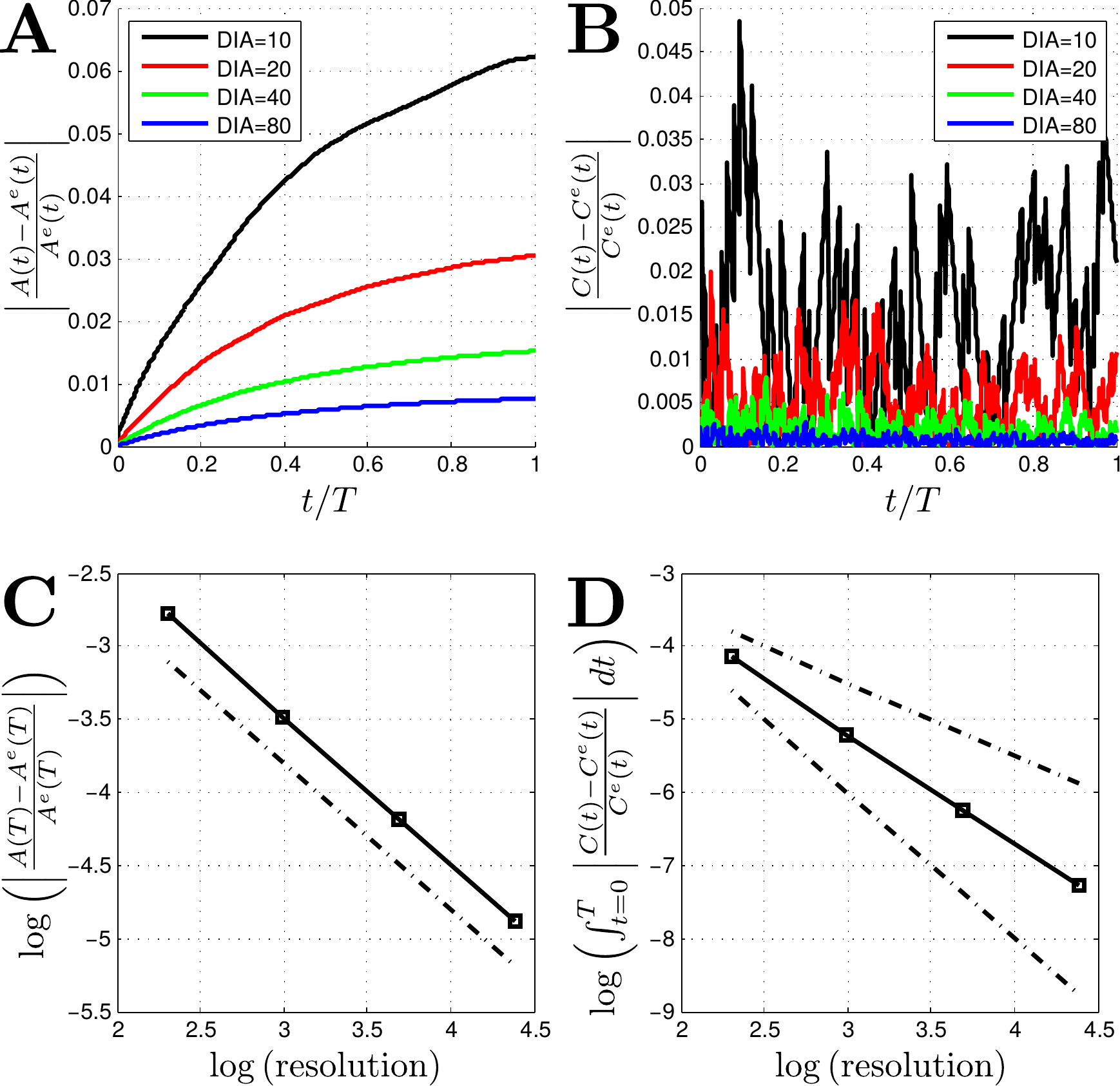}
\par\end{centering}
\caption{\textbf{Case III Validation: Fully Coupled System.}
An initially circular cell is growing through addition of mass.
The mass is added by applying a mass source on the entire cellular domain, 
but the mass source strength is proportional to the concentration of the intracellular concentration.
Therefore, the cellular area should increase linearly in time.
Four different spatial resolutions have been computed, corresponding to a resolution of $\mbox{DIA}=\{10,20,40,80\}$ lattice points for the diameter of the initial circular cell.
{\bf A} The time evolution of the relative error of the cell area for 4 different lattice resolutions.
{\bf B} The time evolution of the relative error of the spatially integrated concentrations (mass conservation) for 4 different lattice resolutions.
{\bf C} Convergence plot of the relative error of the cell area as a function of the lattice resolution.
The fitted lattices is -1.01. The dashed line denotes a slope of -1. 
{\bf D} Convergence plot of the relative error of the integrated concentration (mass conservation) as a function of the lattice resolution.
The fitted slope is 1.50. The dashed lines denote slopes of -1 and -2, respectively.
}
\label{fig:figure_S4}
\end{figure}

The validation results for Case III are given in Figure \ref{fig:figure_S4}.
Figure \ref{fig:figure_S4}{\bf A} shows the relative error temporal evolution of the cell area for the different resolutions,
and \ref{fig:figure_S4}{\bf B} the relative error temporal evolution of the integrated concentration.
The concentration error is much more noisy because the it is only evaluated on lattice points.
When the cell grows, the membrane sweeps across the lattice, and whenever a new lattice site enters the intracellular domain, the 
integrated concentration increases abruptly.
To compute the convergence w.r.t. to spatial and temporal resolution, the relative errors at $T^{ND}=1$ are evaluated for the area, 
but the temporal mean is taken for the concentration convergence analysis.
Figure \ref{fig:figure_S4}{\bf C} shows the convergence analysis for the area.
For the finer grids, the convergence order is approximately 1.
Figure \ref{fig:figure_S4}{\bf D} depicts the convergence analysis for the integrated concentration,
where the fitted order of convergence is $\approx 1.5$.

Finally, the immersed boundary force distribution is validated (Case IV).
The initial setup is similar to the Cases I-III, but a constant membrane force is applied, and there is no mass source.
Due to numerical leakage, the cell starts to shrink and looses mass/area.
The dimensionless force is chosen to be $F^{ND}=1e3$, and the dimension conversion is described in Table \ref{table:forcediscretization}.
The discretization of the validation setup can be found in Table \ref{table:lbdiscretization2}.

\begin{table}[h]\scriptsize
\begin{tabular}{|l|l|l|}
\hline
& ND  & LB  \\
\hline
\hline
mass density & $\rho^{ND} $ & $\rho^{LB} $ \\
\hline
mass units $[M]$ & $[1]$ & $[\delta_{m}^{LB\rightarrow ND}]$ \\
\hline
mass density units 2D $[M/L^{2}]$ & $[1]$ & $\left[ \frac{\delta_{m}^{LB\rightarrow ND}}{\left( \delta_{x}^{LB\rightarrow ND} \right)^{2}}  \right]$\\
\hline
force & $F^{ND}$ & $F^{LB}$ \\
\hline
force units $[ML/T^{2}]$ & $[1]$ & $\left[ \frac{\delta_{m}^{LB\rightarrow ND} \delta_{x}^{LB\rightarrow ND}}{\left( \delta_{t}^{LB\rightarrow ND} \right) ^{2}} \right]$ \\
\hline
\end{tabular}
\caption{Lattice-Boltzmann discretization of mass and force.
ND and LB denote non-dimensional and Lattice Boltzmann units, respectively.
}
\label{table:forcediscretization}
\end{table}

\begin{table}[h]\scriptsize
\begin{tabular}{|l|l|l|l|l|}
\hline
circle diameter & 80 & 160 & 320 & 640 \\
\hline
domain size & 100x100 & 200x200 & 400x400 & 800x800 \\
\hline
$\tau$ fluid & 1 & 1 & 1 & 1 \\
\hline
$\nu^{LB}$ fluid & 1/6 & 1/6 & 1/6 & 1/6 \\
\hline
$\tau$ diff & 1 & 1 & 1 & 1\\
\hline
$\nu^{LB}$ diff & 1/6 & 1/6 & 1/6 & 1/6 \\
\hline
$\delta_{x}$ & 0.1 & 0.05 & 0.025 & 0.0125 \\
\hline
$T^{LB}$ & 1e4 & 4e4 & 1.6e5 & 6.4e5 \\
\hline
$\delta_{t}$ & 1e-4 & 2.5e-5 & 6.25e-6 & 1.5625e-06 \\
\hline
mass density $\rho^{ND}=1$ & 1 & 1 & 1 & 1\\
\hline
$\delta_{m} = \frac{\rho^{ND}}{\rho^{LB}} \left(\delta_{x}\right)^{2} $ & 1e-2 & 2.5e-3 & 6.25e-4 & 1.5625e-4\\
\hline
$F^{LB} = F^{ND}\frac{\delta_{t}^{2}}{\delta_{m} \delta_{x}}$ & 1e-3 & 5e-4 & 2.5e-4 & 1.25e-4\\
\hline
\end{tabular}
\caption{Lattice-Boltzmann discretization of the convergence analysis of Case IV.
ND and LB denote non-dimensional and Lattice Boltzmann units, respectively.}
\label{table:lbdiscretization2}
\end{table}

Besides the spatial and temporal discretization of the lattice, also the discretization of the immersed boundary is studied in Case IV.
The distance between any two neighboring boundary points is shorter than the parameter {\tt MAXLENGTH}, which is always a factor of the 
lattice discretization $\delta_{x}$.
If {\tt MAXLENGTH} is set to 0.5, then the distance between two boundary points does not exceed half of the lattice spacing.
{\tt MAXLENGTH}=0.5 is a frequently used constant (\cite{Peskin1993}); however, here, we also tested smaller (0.1) and larger (2.5) values.

\begin{figure}[t!]
\begin{centering}
\includegraphics[width=\columnwidth]{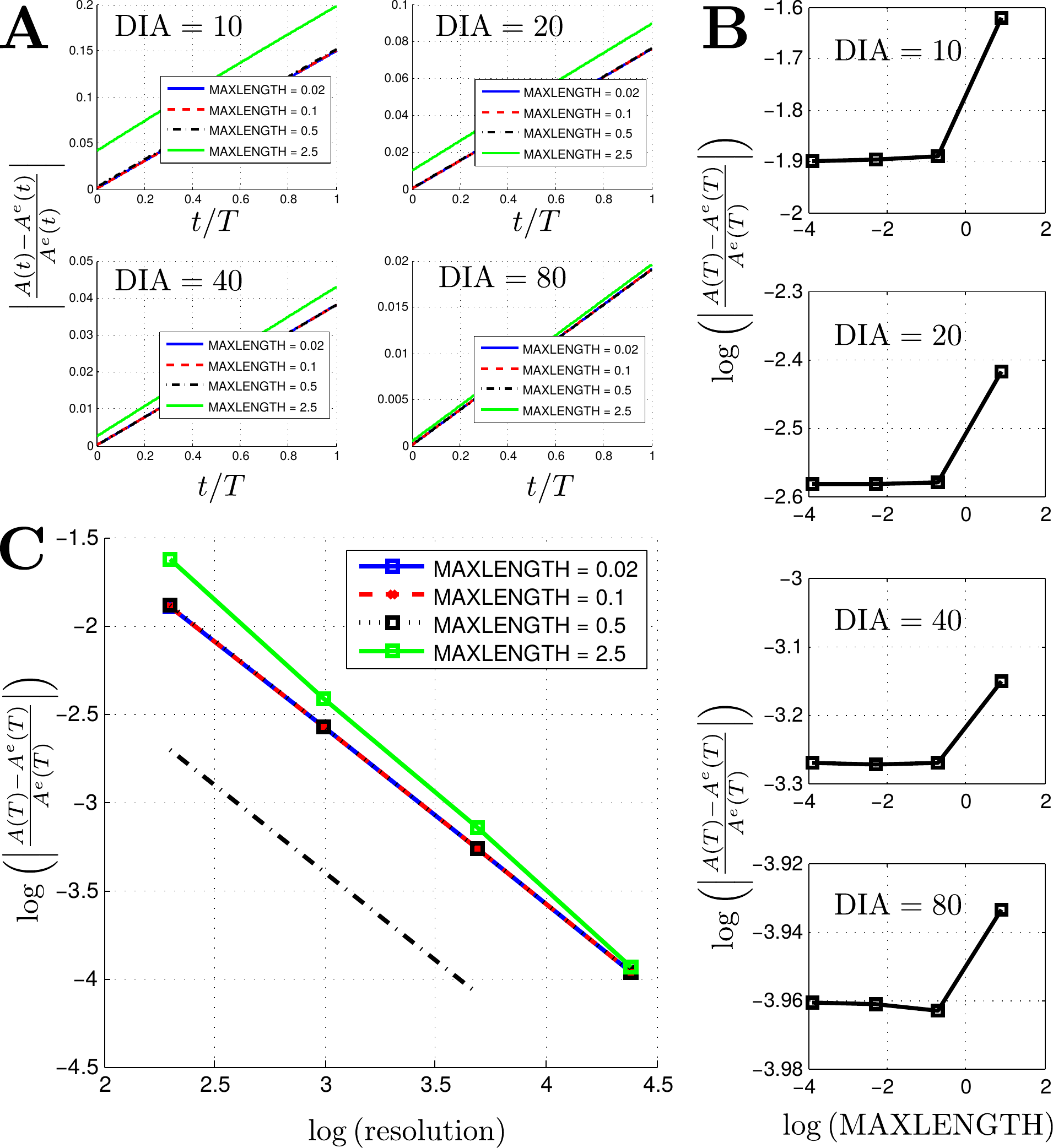}
\par\end{centering}
\caption{\textbf{Case IV Validation: Mass Conservation}
A constant membrane tension is applied to an initially circular cell.
The contractile force will lead to a shrinking cell due to mass leaking of the immersed boundary.
Here, the influence of the immersed boundary discretization is investigated.
Four different spatial resolutions have been computed, corresponding to a resolution of $\mbox{DIA}=\{10,20,40,80\}$ lattice points for the diameter of the initial circular cell.
{\tt MAXLENGTH} denotes the maximally tolerated distance between each two immersed boundary points (measured in space discretization units).
{\bf A} The time evolution of the relative error of the cell area for 4 different lattice resolutions, and for 3 values of {\tt MAXLENGTH}.
{\bf B} The relative error of the cell area as a function of {\tt MAXLENGTH} for different lattice resolutions.
{\bf C} Convergence plot of the relative error of the cell area as a function of the lattice resolution, plotted for 3 values of {\tt MAXLENGTH}.
The fitted slope of the two finest lattices is 0.9891.
}
\label{fig:figure_S5}
\end{figure}

The validation results of Case IV are showed in Figure \ref{fig:figure_S5}.
In Figure \ref{fig:figure_S5}{\bf A}, the evolution of the relative errors of the area are shown for different lattice resolutions.
Figure \ref{fig:figure_S5}{\bf B} shows the convergence as a function of {\tt MAXLENGTH}.
The {\tt MAXLENGTH}$=\{0.1,0.5\}$ values lead to almost similar results, thus confirming that {\tt MAXLENGTH}=0.5 is indeed sufficient.
The convergence as a function of the lattice resolution is shown in Figure \ref{fig:figure_S5}{\bf C}.
The fitted order of convergence is approximately 1.
{\tt MAXLENGTH}$=\{0.1,0.5\}$ lead to indistinguishable results; 
however, even {\tt MAXLENGTH}=2.5 still leads to converging results.

Here, for the convergence analyses, the fluid, mass sources, fluid outlet boundary conditions, the advection and diffusion of a species, its no-flux boundary condition, the immersed boundary and membrane forces are considered.
We showed that all aspects are converging and conclude that the net order of convergence of the fully coupled system is 1.
Depending on the aspect under consideration, it can be higher.
It is well known that the standard Immersed Boundary method suffers from fluid leakage, because the discretized kernel function is not divergence-free.
Several improvements have been proposed (e.g. \cite{Peskin1993,Niu2006}).

%--------------------------------------------------------------------------------------------------------------------------------------------------------
\subsection{Comparison to Farhadifar et al., 2007}\label{Farhadifar}
%--------------------------------------------------------------------------------------------------------------------------------------------------------

The cell topologies are compared to results of the cell vertex model as presented in \cite{Farhadifar2007}.
Starting from a single cell, a tissue with more than 1000 cells is grown with a constant mass source of 0.004 (in LB units).
A cell is divided in a random direction if it exceeds an area of 380 (in LB units).
The membrane tension is varied ($F^{LB}=\{0.001,0.01,0.02\}$ in LB units).

\begin{figure}[t!]
\begin{centering}
\includegraphics[width=\columnwidth]{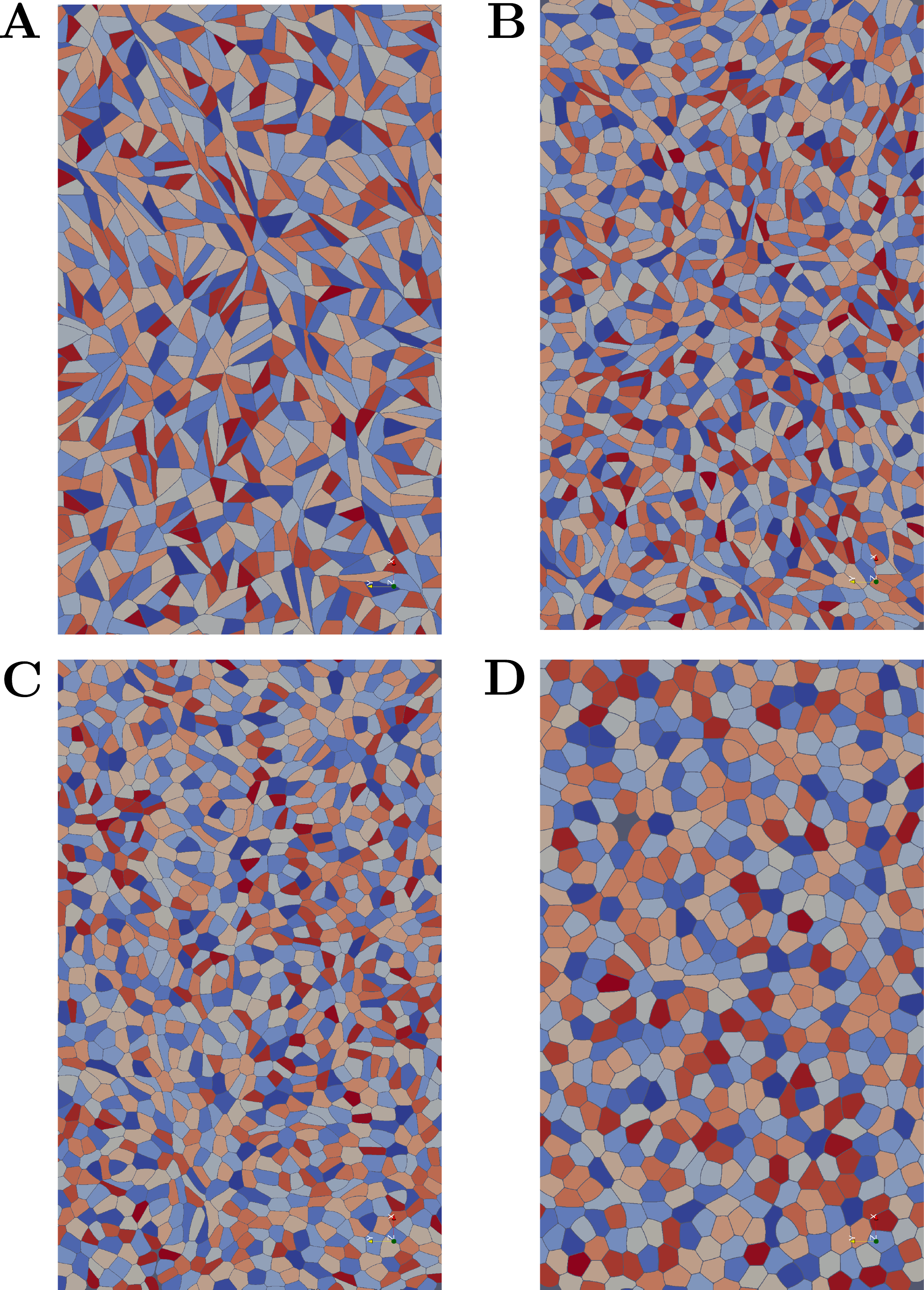}
\par\end{centering}
\caption{\textbf{Cell Topologies for Different Membrane Forces.}
{\bf A} The membrane force constant is small ($F^{LB}=0.001$), such that the cells cannot rearrange adequately.
{\bf B} The membrane force constant is relatively high ($F^{LB}=0.01$), such that the rearrangement activity alters the topology significantly.
{\bf C} The membrane force constant is even higher ($F^{LB}=0.02$).
{\bf D} The tissue shown in {\bf B} is relaxed towards equilibrium.
The color code denotes the domain identifier value (red = young cells, blue = old cells)
}
\label{fig:figure_S6}
\end{figure}

The snapshots in Figure \ref{fig:figure_S6} show the topology for different values of membrane tension ({\bf A} very low membrane tension; {\bf B} medium membrane tension; {\bf C} high membrane tension).
The higher the membrane tension, the higher the rearrangement activity because cell-cell junctions are broken more easily, and
the cells can more easily assume a preferential shape (hexagonal in the domain, circular at the boundary).
In Figure \ref{fig:figure_S6}{\bf D}, the tissue shown in figure \ref{fig:figure_S6}{\bf B} is relaxed, i.e. cell division is stopped,
and only rearrangement is occuring towards an equilibrium configuration.

\begin{figure}[t!]
\begin{centering}
\includegraphics[width=0.8\columnwidth]{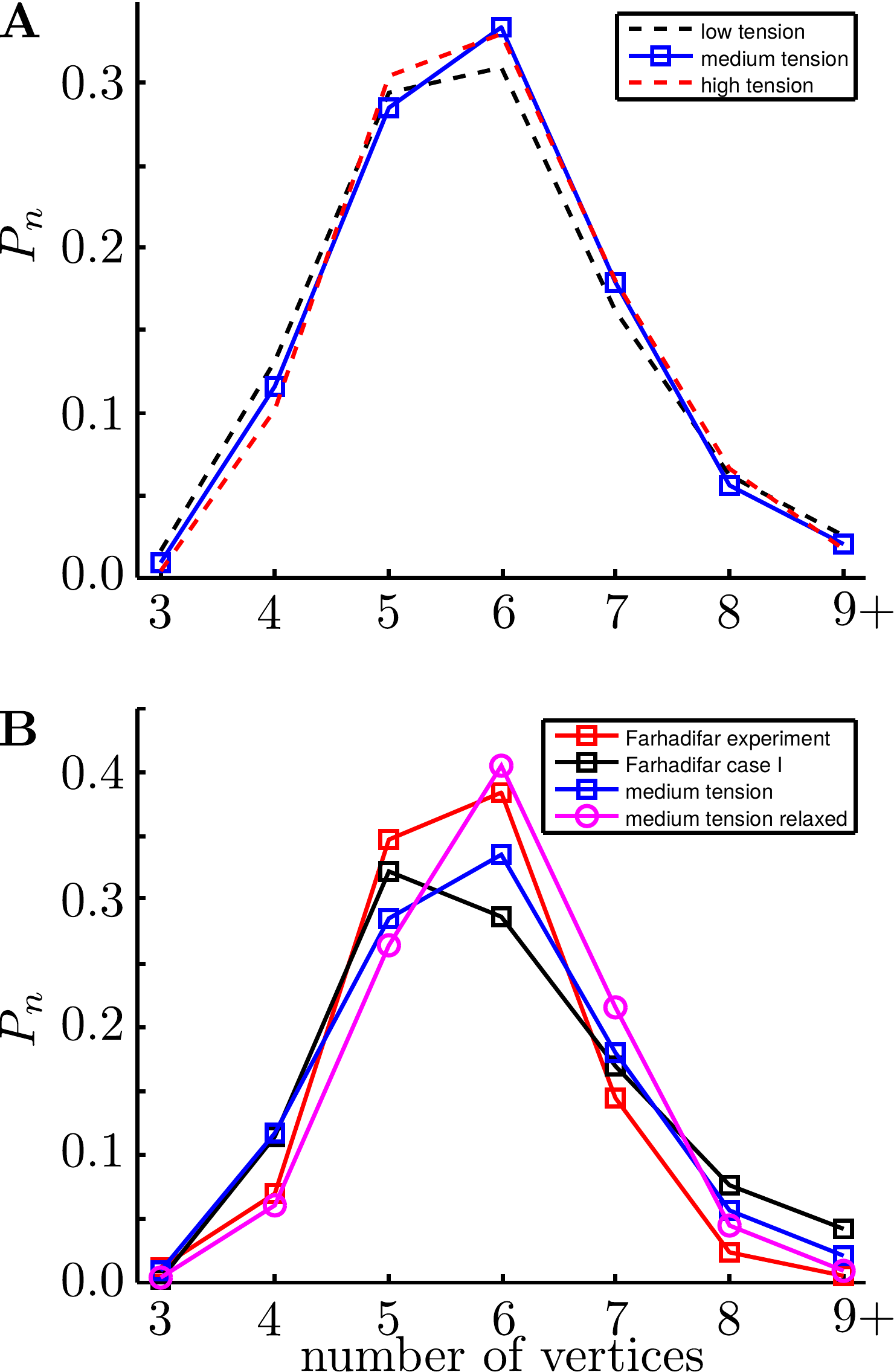}
\par\end{centering}
\caption{\textbf{Comparison of the Cell Topologies to Literature Results}.
$P_{n}$ denotes the occurrence fraction of an $n$-sided polygon. Polygons with $n>9$ are lumped into $n=9$.
{\bf A}
'Low/medium/high tension' correspond to different membrane tensions ($F^{LB}=\{0.001,0.01,0.02\}$ in LB units).
'high tension' has a slightly higher occurrence of penta- and hexagons.
{\bf B}
'Farhadifar experiment' is the polygon distribution in a third instar \textit{Drosophila} wing disc.
'Farhadifar case I' and 'Farhadifar case II' denote simulations with low and high relative contractility, respectively.
'Medium tension relaxed' corresponds to an equilibrated  simulation with $F^{LB}=0.01$ after proliferation has stopped.
}
\label{fig:figure_S7}
\end{figure}

For these cases, the number of vertices are counted.
The result is shown in Figure \ref{fig:figure_S7}, together with the experimental and model results of \cite{Farhadifar2007}.
Although the topologies for different membrane tensions differ significantly, the distribution of the number of vertices (equivalent to the number of neighbors) is not significantly affected.
The relaxed topology, however, is shifted towards higher occurrence of hexagons.
Overall, our results are in agreement with the experimental and simulation results obtained by \cite{Farhadifar2007}.

\end{document}